\documentclass[useAMS,usenatbib]{mn2e}

\usepackage{psfig, epsf, epsfig}

\title[${\rm H_2}$ formation in galaxies]{Formation and evolution of
molecular hydrogen in disk galaxies with different masses and Hubble types}
\author[K. Bekki]
{Kenji Bekki${}^1$\thanks{E-mail:
bekki@cyllene.uwa.edu.au} \\
${}^1$ICRAR M468
The University of Western Australia
35 Stirling Hwy, Crawley
Western Australia 6009, Australia}

\begin{document}

\date{Accepted, Received 2005 February 20; in original form }

\pagerange{\pageref{firstpage}--\pageref{lastpage}} \pubyear{2005}

\maketitle

\label{firstpage}

\begin{abstract}

We investigate the physical properties of molecular hydrogen
(${\rm H_2}$) in isolated and interacting disk 
galaxies with different
masses and Hubble types
by using chemodynamical simulations with ${\rm H_2}$ formation
on  dust grains and dust growth and destruction in interstellar medium
(ISM).
We particularly focus on the dependences of  ${\rm H_2}$ gas mass
fractions ($f_{\rm H_2}$),  spatial distributions of H~{\sc i} and ${\rm H_2}$,
and local ${\rm H_2}$-scaling relations
on initial halo masses ($M_{\rm h}$), baryonic fractions ($f_{\rm bary}$),
gas mass fractions ($f_{\rm g}$), and Hubble types.
The principal results are as follows.
The final $f_{\rm H_2}$ can be larger in disk galaxies
with higher $M_{\rm h}$, $f_{\rm bary}$, and $f_{\rm g}$.
Some low-mass disk models  with $M_{\rm h}$ smaller than 
$10^{10} {\rm M}_{\odot}$ show extremely low $f_{\rm H_2}$ and thus no/little star formation,
even if initial $f_{\rm g}$ is quite large ($>0.9$).
Big galactic bulges can severely suppress the formation of ${\rm H_2}$ from H~{\sc i} on dust grains
whereas strong stellar bars can not only enhance $f_{\rm H_2}$ but also
be responsible for the formation of   ${\rm H_2}$-dominated central rings.
The projected radial distributions of ${\rm H_2}$
are significantly more compact than those of H~{\sc i} and 
the simulated radial profiles of ${\rm H_2}$-to-H~{\sc i}-ratios ($R_{\rm mol}$) follow roughly
$R^{-1.5}$ in MW-type disk models.
Galaxy interaction can significantly increase  $f_{\rm H_2}$ and total ${\rm H_2}$ mass
in disk galaxies.
The local surface mass densities of ${\rm H_2}$ can be correlated with
those of dust in a galaxy. 
The observed correlation between $R_{\rm mol}$ and 
gas pressure ($R_{\rm mol} \propto P_{\rm g}^{0.92}$) can be well reproduced
in the simulated disk galaxies.
\end{abstract}


\section{Introduction}

Formation and evolution processes of molecular hydrogen (${\rm H_2}$)
and interstellar dust in galaxies can be strongly coupled,
because the surface of dust grains can be the major formation
sites of ${\rm H_2}$ (e.g., Gould \& Salpeter 1963; Hollenbach \& Salpeter 1971).
Dust has long been considered to play decisive roles in several aspects of star
and galaxy formation, such as radiative cooling processes in star-forming clouds 
(e.g., Herbst 2001) and  the formation of metal-poor low-mass stars in the early universe
(e.g., Schneider \& Omukai 2010). 
Likewise, ${\rm H_2}$ is an essential element in giant molecular clouds
where star formation is ongoing
(e.g., Blitz et al. 2007; Fukui \& Kawamura 2010)
and its physical properties (e.g., mass densities) are  key parameters for the observed
star-formation laws in galaxies (e.g., Bigiel et al. 2008; Leroy et al. 2008).
Thus, the better understanding of the formation and evolution processes
of {\it both} dust and ${\rm H_2}$ in ISM can lead to the deeper understanding of galaxy
formation and evolution in general.

Physical properties of ${\rm H_2}$ have long been investigated for galaxies
with different Hubble types (e.g., Young \& Scoville 1991, YS91; Boselli et al. 2014),
in different environments (e.g., Leon et al. 1998; Wilson et al. 2009),
and at different redshifts (e.g., Daddi eta l. 2010;
Tacconi et al. 2010; Bauermeister et al. 2013),
and their origins have not been clarified yet.
Recent extensive observational studies
on H~{\sc i} and ${\rm H_2}$ properties of galaxies
and their correlations of galaxy parameters
for a large number of galaxy samples have provided
valuable information on scaling relations of gas and stars
and thus new constraints on galaxy formation and evolution.
(e.g., Catinella et al. 2010; Saintonge et al. 2011). 
Resolved structures and kinematics of giant molecular clouds (GMCs) in nearby 
galaxies such as  M31, M33, and the Large Magellanic Cloud (LMC) have enabled 
astronomers to reveal
the physical factors for the conversion from H~{\sc i} to ${\rm H_2}$
and the possible typical lifetime of GMCs in galaxies
(e.g., Blitz et al. 2007; Fukui \& Kawamura 2010).

In spite of these observational progresses,
a number of key long-standing problems on ${\rm H_2}$ properties of galaxy
have not been resolved yet.
Among them is the origin of the diverse ${\rm H_2}$ properties along 
the Hubble sequence (e.g., YS91).
It is well known that the mass-ratio of ${\rm H_2}$ to H~{\sc i} 
(referred to as $R_{\rm mol}$) is quite 
diverse (more than two orders of magnitudes) for a given Hubble type
and it is systematically higher in early-type disk galaxies (YS91).
This diverse ${\rm H_2}$ properties along the Hubble sequence has been confirmed
in the latest observations with a larger number of galaxy
samples (e.g., Boselli et al. 2014).
It is theoretically unclear what global galaxy parameters (e.g., bulge-to-disk-ratios)
can determine the observed ${\rm H_2}$-to-H~{\sc i}-ratios ($R_{\rm mol}$) in galaxies.

Furthermore,  
one of other unresolved problems is related to the observed 
radial distributions of ${\rm H_2}$ in galaxies 
and their correlations with stellar parameters of galaxies (e.g., Blitz et al. 2007).
Wong \& Blitz (2002) showed that the ratio of ${\rm H_2}$ to H~{\sc i} 
surface density in  a galaxy
depends on the projected distance ($R$) from the galactic center such that
it can be best fit to $R^{-1.5}$ (or $R^{-1}$, see their Fig. 14).
Some early-type spirals (e.g., M31) are observed to have 
intriguing ring-like structures (e.g., YS91).
Recent observations have revealed that
the mass-ratio of ${\rm H_2}$ to stars ($M_{\rm H_2}/M_{\ast}$) 
is quite different between galaxies with different Hubble types and even within
a same type (e.g., Boselli et al. 2014).
These observations have not been explained clearly by previous  theoretical studies
of galaxy formation and evolution, though they would have some profound implications on
galaxy formation.

Recent significant progresses in observational studies of ${\rm H_2}$
properties of galaxies with different types at different $z$ appear
to have triggered extensive theoretical and numerical studies of
of ${\rm H_2}$ formation in
galaxies. 
For example,
recent numerical simulations of galaxy formation and evolution  have incorporated 
the conversion of ${\rm H}_2$ formation 
from H~{\sc i} on dust grains
(e.g., Pelupessy et al. 2006, P06;  Robertson \& Kravtsov 2008, 
Gnedin et al. 2009; Christensen et al. 2012; Kuhlen et al. 2012;
Thompson et al. 2014).
Recent semi-analytic models of galaxy formation
(e.g., Fu et al. 2010; Lagos et al. 2012) adopted
the ${\rm H}_2$ formation model dependent gaseous metallicities and
densities  proposed by
Krumholz, McKee \& Tumlinson (2009)
and thereby investigated the time evolution of ${\rm H}_2$ contents in galaxies.

However, these previous numerical  studies of galaxy formation and evolution
with ${\rm H_2}$ formation 
have not incorporated the evolution of dust (e.g., abundances and masses)
explicitly in their models, and accordingly  could not discuss 
the importance of the {\it joint} evolution of dust and ${\rm H_2}$.
It is ideal for any theoretical studies of ${\rm H_2}$ formation in galaxies
to include the formation and evolution processes
of both dust and ${\rm H_2}$ in a self-consistent manner,
because their evolution can be strongly coupled through star formation.
Given that recent observational studies have revealed a number of important
correlations between dust and ${\rm H_2}$ properties in galaxies 
(e.g., Corbelli et al.  2012), it would be essential for theoretical studies
of ${\rm H_2}$ formation in galaxies to incorporate a model for
the formation and evolution of dust in ISM.

Our previous studies constructed a new chemodynamical model that includes
both the ${\rm H_2}$ formation on dust grains and the formation and destruction
of dust in ISM in a self-consistent manner (Bekki 2013a, 2014; B13a and B14,
respectively,  Yozin \& Bekki 2014a). 
 They therefore could discuss the time evolution of dust
and ${\rm H_2}$ properties in galaxies at their  formation epochs. However,
they did not discuss how the physical properties of ${\rm H_2}$ in galaxies
can possibly depend on their  global parameters such as their total  masses
and Hubble types. Therefore, the following three questions are unresolved:  (i)
what can drive
the observed diversity in ${\rm H_2}$ properties between different galaxies,
(ii) whether and how dust can play a role in controlling ${\rm H_2}$ properties
of galaxies,
and (iii) what is responsible for the observed ${\rm H_2}$-dust correlations
in galaxies.

The purpose of this paper  are two-fold as follows.
First, we investigate  whether the adopted model of ${\rm H_2}$ formation
self-consistently including dust evolution can explain the observed fundamental properties
of ${\rm H_2}$ in galaxies. In this  first investigation,
we focus particularly on the observed correlations between
mass-ratios of ${\rm H_2}$ to H~{\sc i} and other galaxy parameters
(e.g., Wong \& Blitz 2002; Blitz \& Rosolowsy 2006,  Blitz et al. 2007).
Second, we investigate how the physical properties of ${\rm H_2}$ in galaxies
depend on the total halo masses ($M_{\rm h}$),
baryonic mass fractions ($f_{\rm bary}$),  gas mass fractions ($f_{\rm g}$),
Hubble types (e.g., bulge-to-disk-ratios, $f_{\rm b}$),
and the formation redshifts of galaxies by using mainly isolated models
of disk and dwarf galaxies. We also try to understand how galaxy interaction
can influence the time evolution of global ${\rm H_2}$ properties of galaxies
in this second investigation.

The plan of the paper is as follows.
We describe some details of the chemodynamical model with the formation and evolution
of dust and ${\rm H}_2$ adopted by the present study in \S 2.
We present the numerical results
on the physical correlations between ${\rm H_2}$ properties 
and galaxy parameters (e.g., masses)
in disk galaxies in \S 3.
In this section, we also discuss the dependences of the results on the adopted
model parameters.
In \S 4, we discuss the latest observational results on ${\rm H}_2$
properties of galaxies and provides some implications of some key results 
derived in this paper. 
We summarize our  conclusions in \S 5.

In the present, we do not discuss the latest results from $ALMA$ (Atacama Large
Millimeter Array), 
such as the molecular outflow driven by AGN (e.g., Combes et al. 2014)
and total ISM masses probed by dust emission in 107 galaxies from $z=0.2$ to $z=2.5$
(Scoville et al. 2014).
These new observational results will be addressed by our future works with a more
sophisticated chemodynamical model with ${\rm H_2}$ formation.

\begin{table*}
\centering
\begin{minipage}{175mm}
\caption{A list of  physical processes implemented in 
the present simulated code.}
\begin{tabular}{lcl}
{ Physical effects }
& {Inclusion or not \footnote{ $\bigcirc$ and $\times$  in the second
column  mean 
inclusion and non-inclusion of the listed physical effect, respectively}} 
& {Specifications} \\
${\rm H_2}$ formation on dust grains & $\bigcirc$ &  
${\rm H_2}$ formation efficiency dependent on $D$ and $ISRF$ \\
${\rm H_2}$ photodissociation by ISRF  & $\bigcirc$ &  
ISRF locally defined for each gas particle \\
Dust formation  & $\bigcirc$ & 
Formation in SNe and AGB stars \\
Dust growth & $\bigcirc$ & 
Dust growth timescale
dependent on gas properties \\
Dust destruction  & $\bigcirc$ & 
Destruction by  SNe (but not by  hot plasma etc) \\
Size evolution of dust   & $\times$ & \\
Size-dependence on dust composition  & $\times$ & \\
Stellar radiation pressure on dust   & $\times$ & \\
Gas-dust hydrodynamical coupling   & $\times$ & \\
Star formation & $\bigcirc$ & 
${\rm H_2}$-dependent recipe \\
SN feedback effects  & $\bigcirc$ & Both SNII and prompt SNIa are included.\\ 
AGN feedback effects  & $\times$ &  \\
Growth of SMBHs   & $\times$ &  \\
Chemical evolution  & $\bigcirc$ &  
11 elements (e.g., C, N, and O) \\
Chemical enrichment by AGB ejecta & $\bigcirc$ & \\
Metallicity-dependent radiative cooling   & $\bigcirc$ &  
Only for $T_{\rm g}>10^4$ K (no ${\rm H_2}$-dependent cooling for low $T_{\rm g}$) \\
Time-dependent IMF  & $\times$ & 
A fixed canonical IMF is adopted.\\
\end{tabular}
\end{minipage}
\end{table*}

\begin{table*}
\centering
\begin{minipage}{175mm}
\caption{A brief summary for the values of key model parameters.} 
\begin{tabular}{llllllllllll}
{ Model no }
& {$M_{\rm h}$ \footnote{ The total mass of dark matter halo of a
disk galaxy in units of ${\rm M}_{\odot}$.}}  
& {$c$ \footnote{ The `$c$' parameter of the adopted NFW profile for
dark matter halo in a galaxy.}}
& {$M_{\rm d}$ \footnote{ The total disk mass (gas +stars)  of a
disk galaxy in units of ${\rm M}_{\odot}$.}}  
& {$R_{\rm d}$ \footnote{ The stellar disk size  of a
disk galaxy in units of kpc.}}  
& {$f_{\rm g}$ \footnote{ The gas mass fraction ($M_{\rm g}/M_{\rm d}$)  in a
disk galaxy.}}  
& {$f_{\rm bary}$ \footnote{ The baryonic mass fraction 
($M_{\rm d}/M_{\rm h}$) in a
disk galaxy.}}  
& {$f_{\rm b}$ \footnote{ The bulge mass fraction ($M_{\rm b}/M_{\rm d}$)  in a
disk galaxy.}}  
& {[Fe/H]$_0$ \footnote{ The initial mean gas-phase metallicity for the disk of a
galaxy.}}  
& {$z$ \footnote{ The virialization redshift $z$ (or formation redshift) of a
dark halo and this $z$ is used to estimate $r_{\rm vir}$ and $c$. }}  
& {Interaction \footnote{ $\bigcirc$ and $\times$ mean 
inclusion and non-inclusion of galaxy interaction, respectively.}} 
& {Comments} \\
M1 & $10^{12}$ & 10.0 & $6.6 \times 10^{10}$ & $17.5$ & 0.09 & 0.066 & 0.17 & 0.0 & 0 & $\times$ 
& Fiducial MW-type  \\ 
M2 & $10^{12}$ & 10.0 & $6.6 \times 10^{10}$ & $17.5$ & 0.05 & 0.066 & 0.17  & 0.06 &  0 & $\times$ 
&  gas-poor \\ 
M3 & $10^{12}$ & 10.0 & $6.6 \times 10^{10}$ & $17.5$ & 0.27 & 0.066 & 0.17 & $-0.16$ &0 & $\times$ 
&  gas-rich  \\ 
M4 & $10^{12}$ & 10.0 & $3.3 \times 10^{10}$ & $17.5$ & 0.09 & 0.033 & 0.17 & 0.0 & 0 & $\times$ 
&  Lower baryonic fraction \\ 
M5 & $10^{12}$ & 10.0 & $1.7 \times 10^{10}$ & $17.5$ & 0.09 & 0.017 & 0.17 & 0.0 & 0 & $\times$ 
&   \\ 
M6 & $10^{12}$ & 10.0 & $1.7 \times 10^{10}$ & $17.5$ & 0.55 & 0.033 & 0.17 & $-0.44$ & 0 & $\times$ 
&   \\ 
M7 & $10^{12}$ & 10.0 & $6.6 \times 10^{10}$ & $17.5$ & 0.09 & 0.066 & 0.0 & 0.0 & 0 & $\times$ 
&  \\ 
M8 & $10^{12}$ & 10.0 & $6.6 \times 10^{10}$ & $17.5$ & 0.09 & 0.066 & 1.0 & 0.0 & 0 & $\times$ 
&  \\ 
M9 & $10^{12}$ & 10.0 & $6.6 \times 10^{10}$ & $17.5$ & 0.09 & 0.066 & 2.0 & 0.0 & 0 & $\times$ 
&  \\ 
M11 & $10^{12}$ & 10.0 & $1.8 \times 10^{10}$ & $17.5$ & 0.33 & 0.018 & 0.17 & 0.0 & 0 & $\times$ 
&  \\ 
M12 & $10^{12}$ & 10.0 & $1.8 \times 10^{10}$ & $17.5$ & 0.33 & 0.018 & 4.0 & 0.0 & 0 & $\times$ 
&  \\ 
M13 & $10^{12}$ & 10.0 & $6.6 \times 10^{10}$ & $43.8$ & 0.09 & 0.066 & 0.17 & 0.0 & 0 & $\times$ 
& LSB model \\ 
M14 & $10^{12}$ & 4.5 & $6.6 \times 10^{10}$ & $11.3$ & 0.09 & 0.066 & 0.17 & 0.0 & 2 & $\times$ 
& High-$z$ disk \\ 
M15 & $10^{12}$ & 4.5 & $3.3 \times 10^{10}$ & $11.3$ & 0.55 & 0.033 & 0.17 & $-0.44$ & 2 & $\times$ 
& High-$z$ disk \\ 
M16 & $10^{12}$ & 10.0 & $6.6 \times 10^{10}$ & $17.5$ & 0.09 & 0.066 & 0.17 & 0.0 & 0 & $\bigcirc$ 
& Prograde interaction \\ 
M17 & $10^{12}$ & 10.0 & $6.6 \times 10^{10}$ & $17.5$ & 0.09 & 0.066 & 0.17 & 0.0 & 0 & $\bigcirc$ 
& Retrograde interaction \\ 
M18 & $10^{12}$ & 10.0 & $6.6 \times 10^{10}$ & $17.5$ & 0.09 & 0.066 & 0.17 & 0.0 & 0 & $\bigcirc$ 
& Highly inclined disk  \\ 
M19 & $10^{12}$ & 10.0 & $6.6 \times 10^{10}$ & $17.5$ & 0.09 & 0.066 & 0.17 & 0.0 & 0 & $\bigcirc$ 
& More distance encounter \\ 
M20 & $10^{11}$ & 12.6 & $6.6 \times 10^{9}$ & $5.5$ & 0.09 & 0.066 & 0.0 & $-0.46$ & 0 & $\times$ 
& Less massive disk  \\ 
M21 & $10^{11}$ & 12.6 & $6.6 \times 10^{9}$ & $5.5$ & 0.27 & 0.066 & 0.0 & $-0.62$ & 0 & $\times$ 
&  \\ 
M22 & $10^{11}$ & 12.6 & $6.6 \times 10^{9}$ & $5.5$ & 0.55 & 0.066 & 0.0 & $-0.90$ & 0 & $\times$ 
&  \\ 
M23 & $10^{11}$ & 12.6 & $6.6 \times 10^{9}$ & $5.5$ & 0.27 & 0.033 & 0.0 & $-0.62$ & 0 & $\times$ 
&  \\ 
M24 & $10^{11}$ & 4.9 & $3.3 \times 10^{9}$ & $3.5$ & 0.55 & 0.033 & 0.0 & $-0.90$ & 2 & $\times$ 
&  \\ 
M25 & $10^{10}$ & 16.0 & $6.6 \times 10^{8}$ & $1.8$ & 0.09 & 0.066 & 0.0 & $-0.92$ & 0 & $\times$ 
&  Dwarf disk \\ 
M26 & $10^{10}$ & 16.0 & $6.6 \times 10^{8}$ & $1.8$ & 0.27 & 0.066 & 0.0 & $-1.08$ & 0 & $\times$ 
&  \\ 
M27 & $10^{10}$ & 16.0 & $6.6 \times 10^{8}$ & $1.8$ & 0.55 & 0.066 & 0.0 & $-1.36$ & 0 & $\times$ 
&  \\ 
M28 & $10^{10}$ & 16.0 & $3.3 \times 10^{8}$ & $1.8$ & 0.55 & 0.033 & 0.0 & $-1.36$ & 0 & $\times$ 
&  \\ 
M29 & $10^{10}$ & 16.0 & $1.3 \times 10^{8}$ & $1.8$ & 0.55 & 0.013 & 0.0 & $-1.36$ & 0 & $\times$ 
&  \\ 
M30 & $10^{10}$ & 16.0 & $6.6 \times 10^{7}$ & $1.8$ & 0.55 & 0.007 & 0.0 & $-1.36$ & 0 & $\times$ 
&  \\ 
M31 & $3 \times 10^{9}$ & 17.8 & $9.9 \times 10^{7}$ & $1.0$ & 0.55 & 0.033 & 0.0 & $-1.60$ & 0 & $\times$ 
&  Low-mass dwarf \\ 
M32 & $3 \times 10^{9}$ & 17.8 & $9.9 \times 10^{6}$ & $1.0$ & 0.98 & 0.003 & 0.0 & $-2.70$ & 0 & $\times$ 
&  Very gas-rich dwarf\\ 
M33 & $10^{9}$ & 20.0 & $ 1.3 \times 10^{7}$ & $0.6$ & 0.55 & 0.013 & 0.0 & $-1.50$ & 0 & $\times$ 
&  The least massive  dwarf \\ 
M34 & $10^{9}$ & 20.0 & $ 7.2 \times 10^{6}$ & $0.6$ & 0.92 & 0.007 & 0.0 & $-2.90$ &  0 & $\times$ 
\\ 
\end{tabular}
\end{minipage}
\end{table*}


\begin{table*}
\centering
\begin{minipage}{80mm}
\caption{Description of the basic parameter values
for the models of star formation, dust, and  chemical evolution.}
\begin{tabular}{lll}
{Parameters}
& {Adopted values} 
& The standard value adopted in most models \\
{$\tau_0$  \footnote{ The dust growth timescale for a gas density of 1 atom cm$^{-3}$
for the adopted variable dust accretion model.}}
& 0.1, 0.2, 0.4, 2 Gyr  & 0.2 Gyr \\
{$\beta_{\rm d}$ \footnote{ The ratio of dust destruction timescale 
of a gas disk.}}
& 1, 2, 4  & 2 \\
{$f_{\rm dust}$ \footnote{ The initial dust-to-metal-ratio 
to dust growth timescale.}}
& 0.1, 0.4  & 0.4\\
Dust yield  & B13a (fixed) & --\\
Chemical yield  &  T95 for SN,  VG97 for AGB & -- \\
{$\alpha$ \footnote{ The initial metallicity gradient (of the gas disk) 
in units of dex kpc$^{-1}$.}}
& 0, $-0.04$  & 0 \\
{$\rho_{\rm th}$  \footnote{$\rho_{\rm th}$ is the threshold ${\rm H_2}$ gas density for star formation
for each gas particle.}}
& 1 cm$^{-3}$ (fixed) & -- \\
IMF & Kroupa (fixed) &  -- \\
\end{tabular}
\end{minipage}
\end{table*}

\section{The model}

\subsection{A new simulation code}

We adopt our new simulation code recently 
developed in our previous works (B13a and B14) in order
to investigate  spatial and temporal variations of ${\rm H_2}$ in disk galaxies
with different masses and Hubble-types.  
The adopted code can be run on GPU-based machines (clusters), where gravitational
calculations can be done on GPUs whereas other calculations (e.g., star formation
and hydrodynamics) can be done on CPUs.
The code adopts
the smoothed-particle hydrodynamics (SPH) method for following the time
evolution of gas dynamics in galaxies.

The present simulations include
the formation of dust grains in the stellar winds of supernovae (SNe)
and asymptotic giant branch (AGB) stars, 
the time evolution of interstellar radiation field (ISRF),
the growth and destruction processes
of dust in the interstellar
medium (ISM),  
the ${\rm H_2}$ formation on dust grains,
and the ${\rm H_2}$ photo-dissociation due to far ultra-violet (FUV)
light in a self-consistent manner.
Although we can investigate 
the formation of polycyclic aromatic hydrocarbon (PAH) dust
in carbon-rich AGB stars and the time evolution of dust by using the simulation code,
we do not extensively investigate the dust properties  in the present study.
Instead, we briefly discuss the possible physical correlations between
dust and ${\rm H_2}$ properties in disk galaxies.

The new simulation code does not include 
the effects of feedback of active galactic nuclei (AGN)
on ISM and the growth of supermassive black holes (SMBHs) in galaxies
so that we can not investigate how the feedback can change the spatial
distributions and mass budgets of dust and ${\rm H_2}$ in galaxies. Although
it is an important issue whether and how the AGN feedback effects can influence
the dust properties of ISM and thus the ${\rm H_2}$ contents, we will
discuss this in our future papers. The new code allows us to choose
whether some physical effects (e.g., dust formation etc)
are included or excluded (i.e., 'switched on or off')
so that we can investigate
how the physical effects are important in the evolution of gas and stars in galaxies.
Accordingly, we summarize the physical effects that are included 
in the present study in Table 1 for clarity.

Since the major formation site of ${\rm H_2}$ in ISM is the surface of dust
grains,  it is crucial for any theoretical study on ${\rm H_2}$ evolution
of galaxies to properly investigate
the time evolution of dust abundances in disk galaxies.
Given that both observational and theoretical studies
have shown that dust-to-metal-ratios can be diverse (e.g., Hirashita 1999; Galametz et al. 2011),
we need to refrain from deriving dust abundances from metallicities by
assuming constant dust-to-metal-ratios.
The present code indeed enables us to investigate the time evolution of
different dust components in galaxies so that we can more properly
predict the time evolution of ${\rm H_2}$ in galaxies.
We mainly focus on  isolated  disk  galaxies composed of  dark matter halo,
stellar disk,  stellar bulge, and  gaseous disk:
Some new results on ${\rm H_2}$ properties in forming galaxies
at high redshifts ($z$) are discussed in B14.
Since the details of
the physical models for the formation and evolution of dust and ${\rm H_2}$
are given in B13 and B14,
we briefly describe the models in the present paper.

\subsection{A disk galaxy}

\subsubsection{Structure and kinematics}

The total masses of dark matter halo, stellar disk, gas disk, and
bulge of a disk galaxy are denoted as $M_{\rm h}$, $M_{\rm s}$, $M_{\rm g}$,
and $M_{\rm b}$, respectively. The total disk mass (gas + stars)
and gas mass fraction
are denoted as $M_{\rm d}$ and $f_{\rm g}$, respectively,  for convenience.
The mass ratio of 
the disk ($M_{\rm s}+M_{\rm g}$)
to the dark matter halo ($M_{\rm h}$)
in a disk galaxy is a `baryonic mass fraction' and denoted as $f_{\rm bary}$. 
The bulge-to-disk-ratio is defined as $M_{\rm b}/M_{\rm s}$ and represented
by a parameter $f_{\rm b}$. The four key parameters in the present study
are $M_{\rm h}$, $f_{\rm g}$, $f_{\rm bary}$, and $f_{\rm b}$.

In order to describe the initial density profile of dark matter halo
in a disk galaxy,
we adopt the density distribution of the NFW
halo (Navarro, Frenk \& White 1996) suggested from CDM simulations:
\begin{equation}
{\rho}(r)=\frac{\rho_{0}}{(r/r_{\rm s})(1+r/r_{\rm s})^2},
\end{equation}
where  $r$, $\rho_{0}$, and $r_{\rm s}$ are
the spherical radius,  the characteristic  density of a dark halo,  and the
scale
length of the halo, respectively.
The $c$-parameter ($c=r_{\rm vir}/r_{\rm s}$, where $r_{\rm vir}$ is the virial
radius of a dark matter halo) and $r_{\rm vir}$ are chosen appropriately
for a given dark halo mass ($M_{\rm dm}$)
by using the $c-M_{\rm h}$ relation for $z=0$ and 2
predicted by recent cosmological simulations 
(e.g., Neto et al. 2007; Mu${\rm \tilde n}$oz-Cuartas et al. 2011).

The bulge of a disk galaxy  has a size of $R_{\rm b}$
and a scale-length of $R_{\rm 0, b}$
and is represented by the Hernquist
density profile. The bulge is assumed to have isotropic velocity dispersion
and the radial velocity dispersion is given according to the Jeans equation
for a spherical system.
The bulge-to-disk ratio ($f_{\rm b}=M_{\rm b}/M_{\rm d}$) of a disk galaxy is 
a free parameter ranging from 0 (pure disk galaxy) to 4 (bulge-dominated).
The `MW-type'  models are those
with $f_{\rm b}=0.17$ and $R_{\rm b}=0.2R_{\rm s}$,
where $R_{\rm s}$ is the stellar disk size of a galaxy.
We adopt the mass-size scaling relation of $R_{\rm b} = C_{\rm b} M_{\rm b}^{0.5}$
for bulges so that we can determine $R_{\rm b}$ for a given $M_{\rm b}$.
The value of $C_{\rm b}$ is determined so that $R_{\rm b}$ can be 3.5 kpc
for $M_{\rm b}=10^{10} {\rm M}_{\odot}$ (corresponding to the mass and size
of the MW's bulge).

The radial ($R$) and vertical ($Z$) density profiles of the stellar disk are
assumed to be proportional to $\exp (-R/R_{0}) $ with scale
length $R_{0} = 0.2R_{\rm s}$  and to ${\rm sech}^2 (Z/Z_{0})$ with scale
length $Z_{0} = 0.04R_{\rm s}$, respectively.
The gas disk with a size  $R_{\rm g}=2R_{\rm s}$
has the  radial and vertical scale lengths
of $0.2R_{\rm g}$ and $0.02R_{\rm g}$, respectively.
In the present model for the MW-type,  the exponential disk
has $R_{\rm s}=17.5$ kpc and
$R_{\rm g}=35$ kpc. 
In addition to the
rotational velocity caused by the gravitational field of disk,
bulge, and dark halo components, the initial radial and azimuthal
velocity dispersions are assigned to the disc component according to
the epicyclic theory with Toomre's parameter $Q$ = 1.5.
The vertical velocity dispersion at a given radius is set to be 0.5
times as large as the radial velocity dispersion at that point.

The total numbers of particles used for dark matter halo ($N_{\rm dm}$), stellar disk ($N_{\rm s}$),
and gaseous disk ($N_{\rm g}$)  in a simulation are 700000, 200000, and 100000 respectively.
The total number of particle for bulge is $f_{\rm b} N_{\rm s}$, which means that 
$N_{\rm b}=33400$ for the MW-type disk galaxy model with $f_{\rm b}=0.167$
and $400000$ for the big bulge model with
$f_{\rm b}=2$. Therefore, the total number of particles
is $N=1033400$ for the fiducial MW-type model and $N=1400000$ for the big bulge model.
The gravitational softening length for each component is determined by the number
of particle used for each component and by the size of the distribution
(e.g., $R_{\rm s}$ and $r_{\rm vir}$)  and the value is described later.

\subsubsection{Gas-phase metallicity and its radial gradient}

Observations have shown that (i) there is a mass-metallicity relation in disk galaxies
(e.g., Tremonti et al. 2004) and (ii) most of disk galaxies show negative metallicity
gradients (i.e., higher metallicity in inner regions) with the slopes being different
in different galaxies (e.g., Zaritsky et al. 1994). We therefore determine the initial
mean metallicity of the gas disk in a galaxy according to the observe mass-metallicity
relation for the adopted total halo  mass ($M_{\rm h}$)
and gas mass fraction for the galaxy. For example,
a MW-type disk galaxy with $M_{\rm h}=10^{12} {\rm M}_{\odot}$
($M_{\rm s}=6 \times 10^{10} {\rm M}_{\odot}$)  and $f_{\rm g}=0.09$
(corresponding to the fiducial model later described) can  have [Fe/H]=0.

The gas-phase  metallicity of each (gaseous and stellar) particle is given
according to its initial position:
at $r$ = $R$,
where $r$ ($R$) is the projected distance (in units of kpc)
from the center of the disk, the metallicity of the star is given as:
\begin{equation}
{\rm [Fe/H]}_{\rm r=R} = {\rm [Fe/H]}_{\rm d, r=0} + \alpha \times {\rm R}. \,
\end{equation}
where  $\alpha$ is the slope of the metallicity gradient in units of dex kpc$^{-1}$.
Since the present results on global properties of ${\rm H_2}$ do not depend on
$\alpha$, we mainly show the results of the models with $\alpha=0$.
We however discuss briefly how the present results can change
if we adopt a steeper metallicity gradient of $\alpha=-0.04$,
which is  the observed  value  of our Milky Way
(e.g., Andrievsky et al. 2004).

\subsubsection{Low-$z$ vs high-$z$ disks}

Since we mainly investigate ${\rm H_2}$ properties of disk galaxies  at $z=0$,
we construct the disk model by  using the $c-M_{\rm h}$ relation for dark matter
halos at  $z=0$
(e.g., Neto et al. 2007; Mu${\rm \tilde n}$oz-Cuartas et al. 2011). 
We however investigate a number of disk models that can represent disk galaxies at high $z$
by using the $c-M_{\rm h}$ relation dependent on $z$ (Mu${\rm \tilde n}$oz-Cuartas et al. 2011).
In constructing the disk models at high $z$, we consider the following two theoretical
and observational results. First,
the mean density of a dark matter halo ($\rho_{\rm dm}$)
with $M_{\rm h}$ at $z$ is $(1+z)^3$ times higher
than that (of a dark matter halo with the same $M_{\rm h}$) at $z=0$ ($\rho_{\rm dm,0}$).
Second, the stellar disk size of a galaxy at $z$ ($R_{\rm s}$)  is $(1+z)^{-0.4}$ times
larger than that at $z=0$ ($R_{\rm s,0}$). 
Therefore the mean density of a dark matter halo  in a disk galaxy at $z=0$ is
as follows; 
\begin{equation}
\rho_{\rm dm}(z)=\rho_{\rm dm,0} (1+z)^3,
\end{equation}
where $r_{\rm vir}(z) \propto \rho_{\rm dm}^{-1/3}$ for a given $M_{\rm h}$.
The stellar disk size is given as follows:
\begin{equation}
R_{\rm s}(z)=R_{\rm s,0} (1+z)^{-0.4},
\end{equation}
which means that the size ratio of $r_{\rm vir}$ to $R_{\rm s}$ is proportional
to $(1+z)^{-0.6}$.
For example, $c$ and $r_{\rm vir}$ for a disk galaxy with
$M_{\rm h}=10^{12} {\rm M}_{\odot}$  at $z=0$ are 10 and 245kpc,
respectively, whereas they are 4.5 and 82 kpc at $z=2$.
The simulated dark matter halos are assumed to have a scaling relation of 
$r_{\rm vir} \propto M_{\rm h}^{0.5}$
at $z=0$ and 2.

\begin{figure*}
\psfig{file=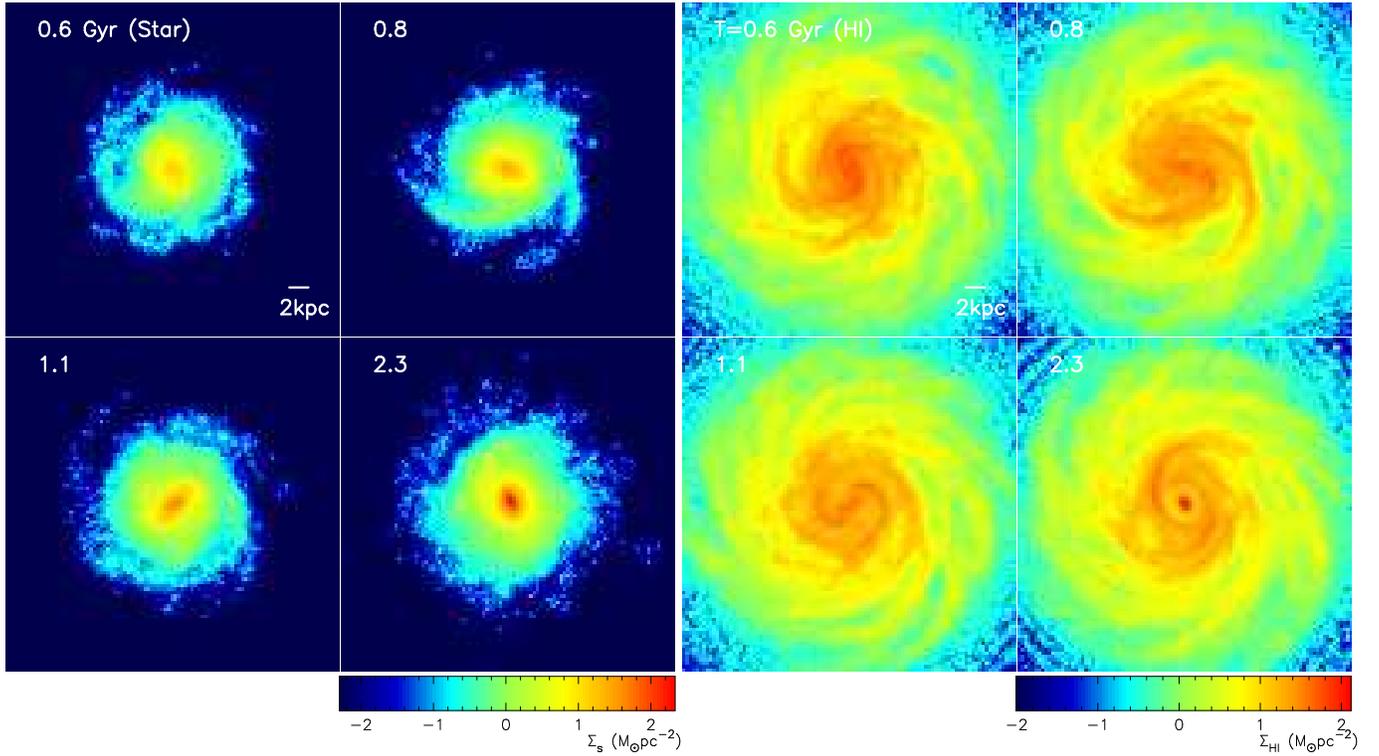,width=18.0cm}
\caption{
The time evolution of the projected mass densities for stars ($\Sigma_{\rm s}$, left four)
and H~{\sc i} ($\Sigma_{HI}$, right four) for the fiducial MW-type disk model (M1). The time $T$,
which represents the time that has elapsed since
the simulation started, is shown in the upper left corner for each panel.
The mass densities are derived from the distributions of particles projected onto the $x$-$y$ plane of
the disk. The projected density fields are smoothed by using a Gaussian kernel with a smoothing length
of 875pc.
}
\label{Figure. 1}
\end{figure*}

\begin{figure*}
\psfig{file=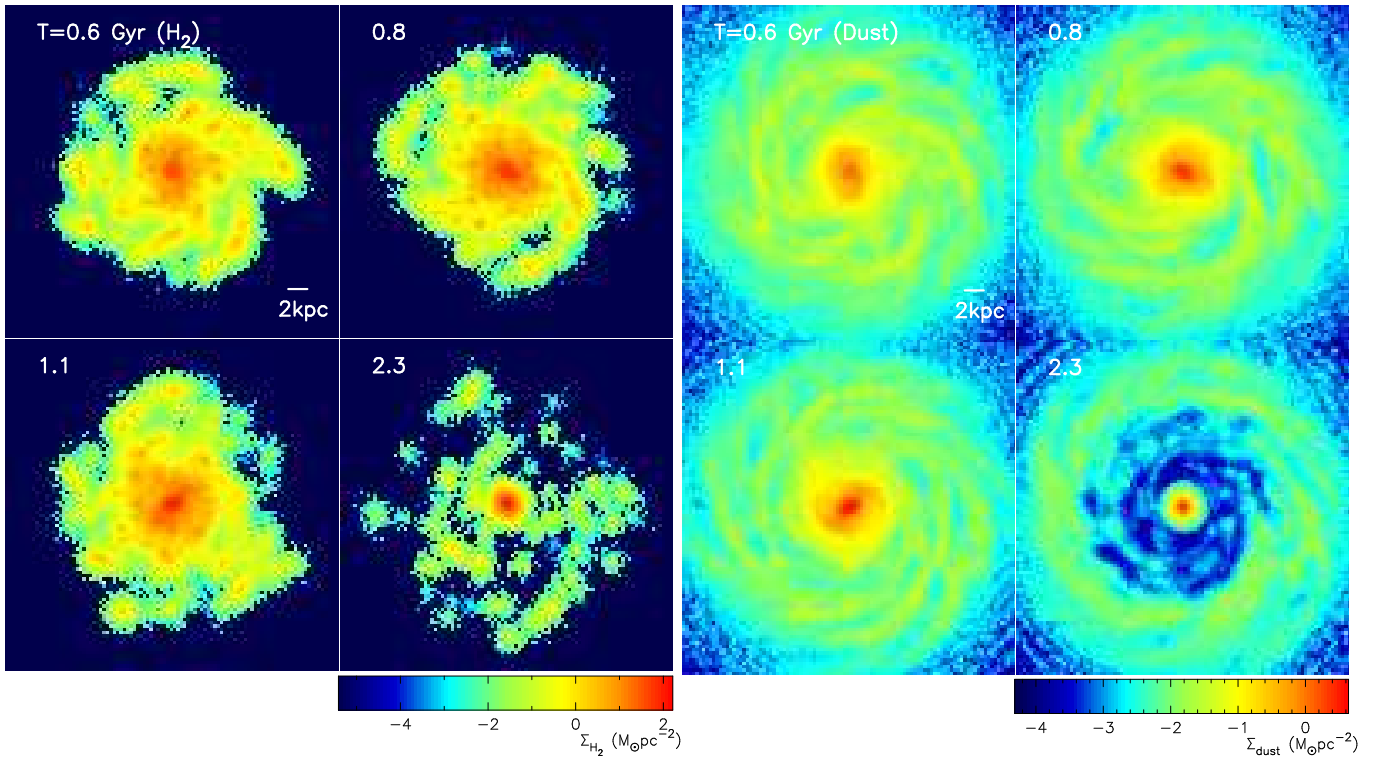,width=18.0cm}
\caption{
The same as Fig. 1 but for ${\rm H_2}$ ($\Sigma_{\rm H_2}$, left four)
and for dust ($\Sigma_{\rm dust}$, right four).
}
\label{Figure. 2}
\end{figure*}

\begin{figure*}
\psfig{file=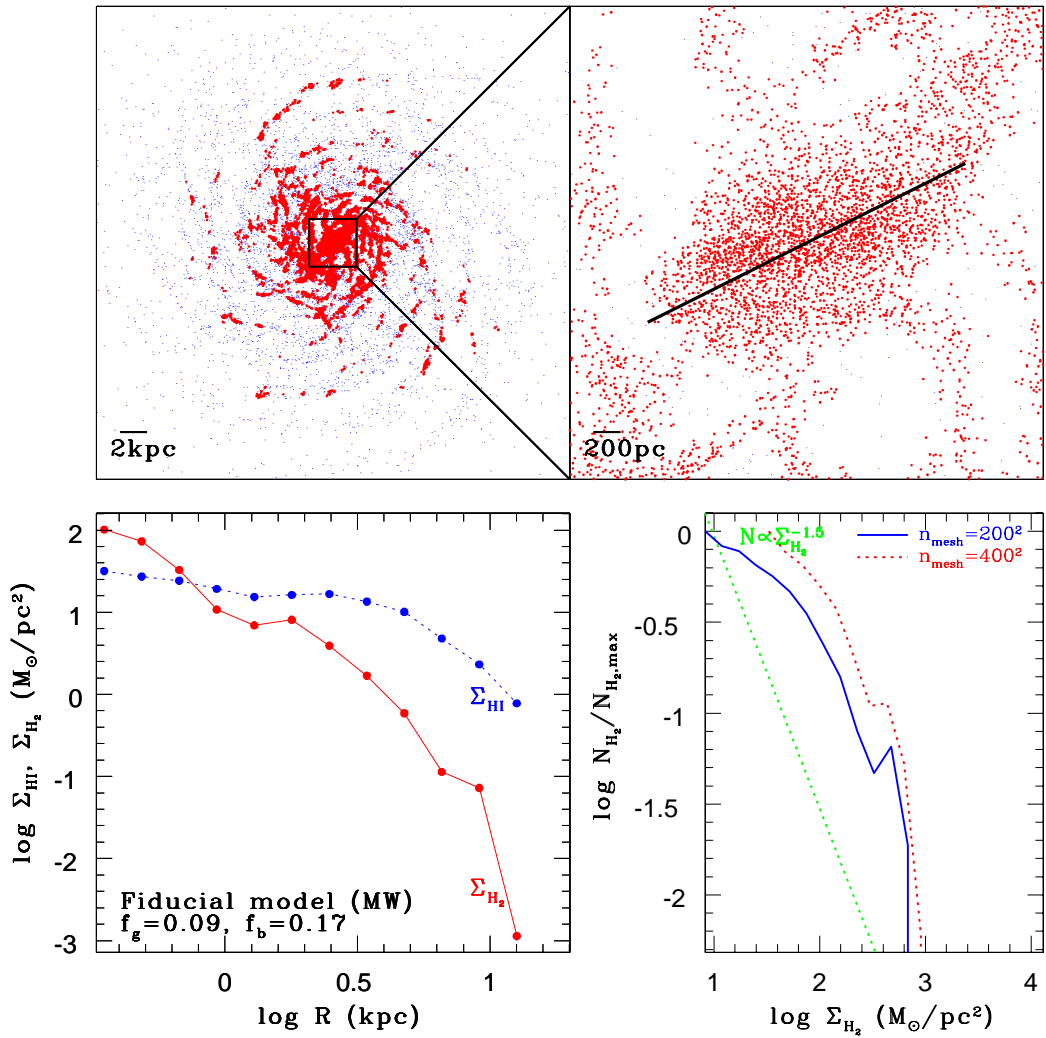,width=14.0cm}
\caption{
The upper two panels show the distributions of gas with $f_{\rm H_2}<0.01$ 
(blue, corresponding to H~{\sc i}-dominated gas)
and $f_{\rm H_2} \ge 0.01$ (red)
projected onto the $x$-$y$ plane at $T=1.1$ Gyr in
the fiducial MW-type disk model M1. The physical scales in these two panels are different
and indicated by a thick bar in each panel. 
The blue and red dots therefore can represent H~{\sc i} (gas with no/little ${\rm H_2}$)
and ${\rm H_2}$,
respectively.
The central black thick line in the upper right panel indicates the location of the central
stellar bar in this model.
Only one in ten particles is shown so that the file
size of this figure (in .eps format) can be  small enough  (yet informative to readers) in these upper
panels.
The lower left panels shows the (projected) radial distribution of H~{\sc i} ($\Sigma_{\rm HI}$, blue dotted)
and ${\rm H_2}$ ($\Sigma_{\rm H_2}$, red solid) in the fiducial model.
The lower right panel shows the ${\rm H_2}$ surface density distribution (H2SDD) 
normalized by the maximum number of $N_{\rm H_2}$ ($N_{\rm H_2,max}$,
where $N_{\rm H_2}$ is the number of meshes with a given ${\rm H_2}$ surface density,
$\Sigma_{\rm H_2}$)  among  local regions
for $n_{\rm mesh}=200^2$ (blue solid) and $n_{\rm mesh}=400^2$ (red dotted)
in the fiducial model. The dotted green line indicates
$N\propto \Sigma_{\rm H_2}^{-1.5}$,  which is similar to the observed
molecular cloud mass function (MCMF) of the Galaxy that can be approximated
as $N\propto m_{\rm mc}^{-1.5}$, where $m_{\rm mc}$ is the mass of a molecular cloud.
}
\label{Figure. 3}
\end{figure*}

\begin{figure*}
\psfig{file=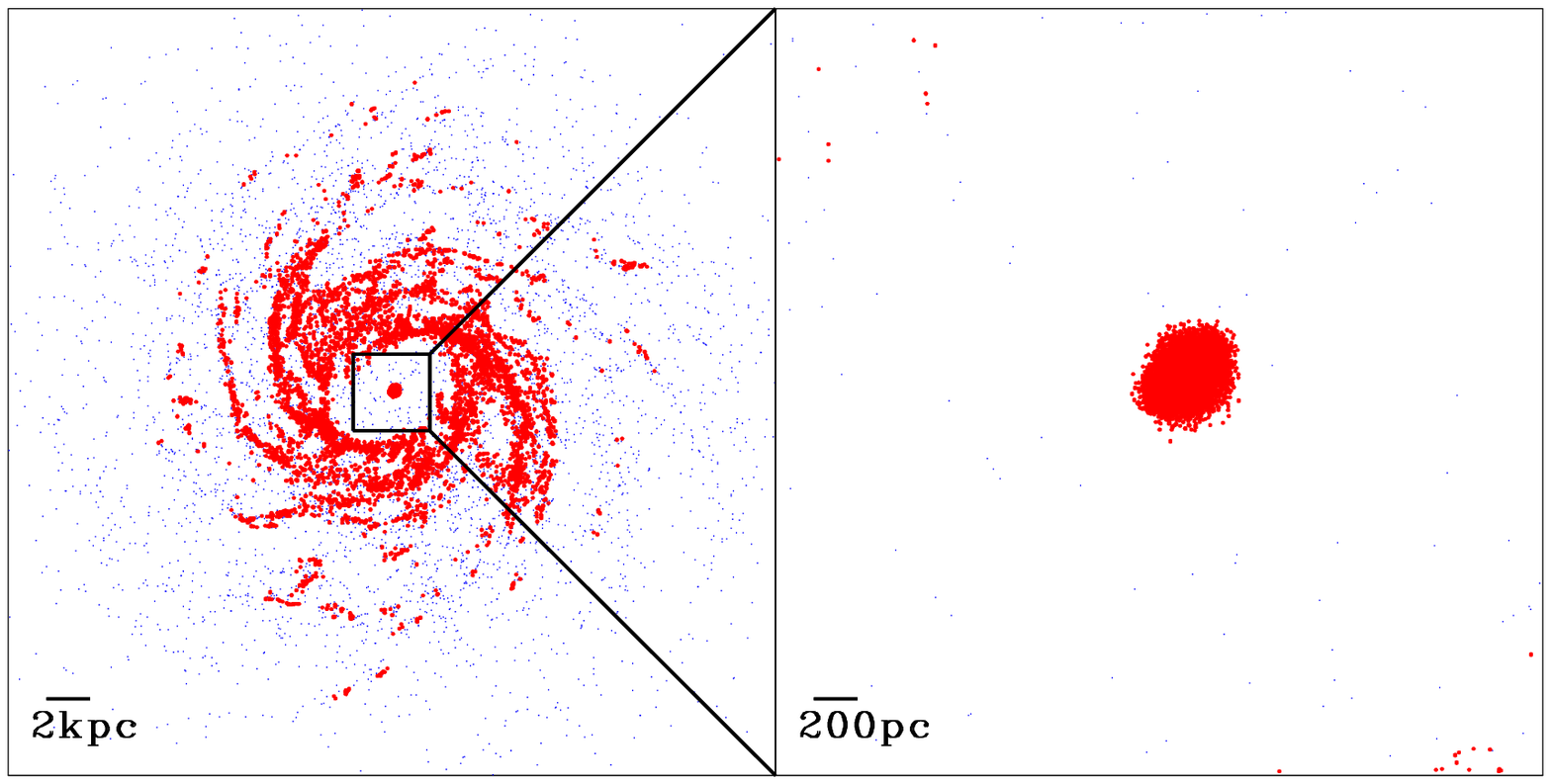,width=14.0cm}
\caption{
The distributions of gas with $f_{\rm H_2}<0.01$ (blue)
and $f_{\rm H_2} \ge 0.01$ (red) projected onto the $x$-$y$ plane at $T=2.3$ Gyr in
the fiducial MW-type disk model M1. 
}
\label{Figure. 4}
\end{figure*}

\subsection{Star formation}

Since SF can proceed in molecular clouds,
we adopt the following `${\rm H_2}$-dependent' SF recipe
(B13a) using molecular gas fraction
($f_{\rm H_2}$) defined for each gas particle in the present study.
A gas particle {\it can be} converted
into a new star if (i) the local dynamical time scale is shorter
than the sound crossing time scale (mimicking
the Jeans instability) , (ii) the local velocity
field is identified as being consistent with gravitationally collapsing
(i.e., div {\bf v}$<0$),
and (iii) the local density exceeds a threshold density for star formation ($\rho_{\rm th}$).
We mainly investigate the models with $\rho_{\rm th}=1$ cm$^{-3}$
in the present study.

A gas particle can be regarded as a `SF candidate' gas particle
if the above three SF conditions (i)-(iii) are satisfied.
It could be possible to convert some fraction ($\propto f_{\rm H_2}$)
of a SF candidate  gas particle
into a new star at each time step until the mass of the gas particle
becomes very small. However, this SF conversion method can increase dramatically
the total number of stellar particles, which becomes  numerically very costly.
We therefore adopt the following SF conversion method.
A SF candidate $i$-th gas
particle is regarded as having  a SF probability ($P_{\rm sf}$);
\begin{equation}
P_{\rm sf}=1-\exp ( -C_{\rm eff} f_{\rm H_2}
\Delta t {\rho}^{\alpha_{\rm sf}-1} ),
\end{equation}
where $C_{\rm eff}$ corresponds to a star formation  efficiency (SFE)
in molecular cores and is set to be 1,
$\Delta t$ is the time step width for the gas particle,
$\rho$ is the gas density of the particle,
and $\alpha_{\rm sf}$ is
the power-law slope of the  Kennicutt-Schmidt law
(SFR$\propto \rho_{\rm g}^{\alpha_{\rm sf}}$;  Kennicutt 1998).
A reasonable value of
$\alpha_{\rm sf}=1.5$ is adopted in the present
study.
This SF probability has been already introduced in our early chemodynamical
simulations of galaxies (e.g., Bekki \& Shioya 1998).

At each time step   random numbers ($R_{\rm sf}$; $0\le R_{\rm sf}  \le 1$)
are generated and compared with $P_{\rm sf}$.
If $R_{\rm sf} < P_{\rm sf}$, then the gas particle can be converted into
a new stellar one.
In this SF recipe, a gas particle with a higher gas density
and thus a shorter SF timescale ($\propto
\rho/\dot{\rho} \propto \rho^{1-\alpha_{\rm sf}}$)
can be more rapidly converted into a new star owing to the larger
$P_{\rm sf}$. Equally, a gas particle with a higher $f_{\rm H_2}$
can be more rapidly converted into a new star.
We thus consider that the present SF model is a good approximation
for star formation in molecular gas of disk galaxies.

Each SN is assumed to eject the feedback energy ($E_{\rm sn}$)
of $10^{51}$ erg and 90\% and 10\% of $E_{\rm sn}$ are used for the increase
of thermal energy (`thermal feedback')
and random motion (`kinematic feedback'), respectively.
The thermal energy is used for the `adiabatic expansion phase', where each SN can remain
adiabatic for a timescale of $t_{\rm adi}$.
Although $t_{\rm adi}=10^5$ yr is reasonable for a single SN explosion,
we adopt a much longer $t_{\rm adi}$ of $\sim 10^6$ yr.
This is mainly because multiple  SN explosions can occur for a gas particle with a mass of
$10^5 {\rm M}_{\odot}$ in these galaxy-scale simulations,
and $t_{\rm adi}$ can be different
for multiple SN explosions in a small local region owing to complicated
interaction between gaseous ejecta from different SNe.
Such interaction of multiple  SN explosions would make the adiabatic phase
significantly longer in real ISM of galaxies.

\subsection{IMF}

We adopt a canonical stellar initial mass function (IMF) proposed by Kroupa (2001), which
has three different slopes at different mass ranges. Our recent study (Bekki 2013b, B13b)
has shown that if the IMF depends on local physical properties of ISM (e.g., gas density and metallicity),
then galaxy evolution, in particular, the time evolution of ${\rm H_2}$ and dust contents,
can be significantly influenced by the time-varying IMF.
Since the main purpose of this study is not to demonstrate how the ${\rm H_2}$ contents of 
disk galaxies can be influenced by a time-varying IMF,
we assume that the IMF is fixed
at the adopted Kroupa-IMF in the present study.
It is our future works to clarify the roles of time-varying IMFs in controlling ${\rm H_2}$ properties
in galaxies with different masses and Hubble types.

\begin{figure*}
\psfig{file=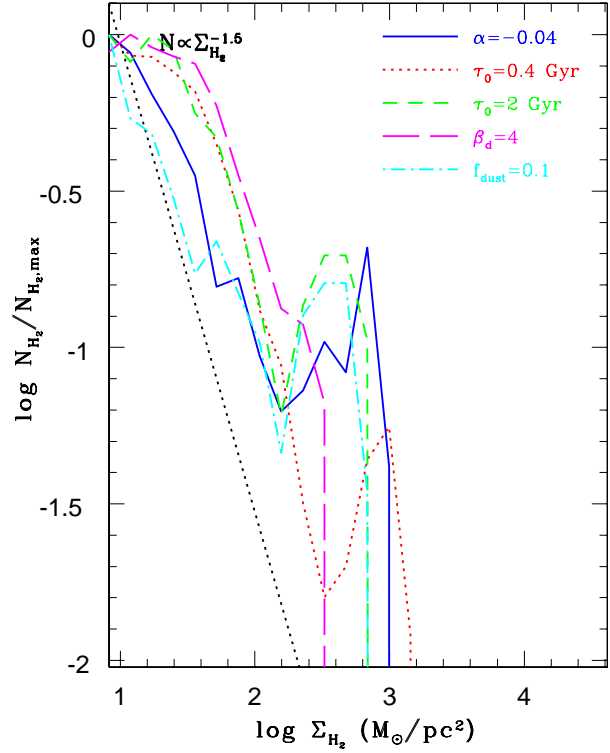,width=8.0cm}
\caption{
The H2SDD (${\rm H_2}$ surface density distribution)
in different five MW-type disk  models (M1): $\alpha=-0.04$ (blue solid), $\tau_0=0.4$ Gyr (red dotted),
$\tau_0=2$ Gyr (green short-dashed), $\beta_{\rm d}=4$ (magenta long-dashed), 
and $f_{\rm dust}=0.1$ (cyan, dot-dashed).
}
\label{Figure. 5}
\end{figure*}

\subsection{Evolution of dust and metals}

\subsubsection{Chemical enrichment}

Since the present model for chemical enrichment processes of galaxies
is exactly the same as that used in B13a, we briefly describe the
model here.
Chemical enrichment through star formation and metal ejection from
SNIa, II, and AGB stars is self-consistently included in the chemodynamical
simulations.
We investigate the time evolution of the 11 chemical elements of H, He, C, N, O, Fe,
Mg, Ca, Si, S, and Ba in order to predict both chemical abundances and dust properties
in the present study. 
We consider the time delay between the epoch of star formation
and those  of supernova explosions and commencement of AGB phases (i.e.,
non-instantaneous recycling of chemical elements).

We adopt the `prompt SN Ia' model in which
the delay time distribution (DTD)
of SNe Ia is consistent with  recent observational results by  extensive SN Ia surveys
(e.g.,  Mannucci et al. 2006).
In this prompt SN Ia mode,
there is a time delay ($t_{\rm Ia}$) between the star formation
and the metal ejection for SNe Ia and the range of $t_{\rm Ia}$ is
0.1 Gyr $\le t_{\rm Ia} \le$ 10 Gyr.
The fraction of the stars that eventually
produce SNe Ia for 3--8$M_{\odot}$ has not been observationally determined.
In  the present study,  $f_{\rm b}=0.05$ is assumed.
We adopt the nucleosynthesis yields of SNe II and Ia from Tsujimoto et al. (1995; T95)
and AGB stars from van den Hoek \& Groenewegen (1997; VG97)
in order to estimate chemical yields in the present study.

\subsubsection{Dust model}

Since the dust model adopted in the present study is the same as those in B13a and B14,
we here briefly describe the model.
We calculate the  total mass of $j$th component ($j$=C, O, Mg, Si, S, Ca, and Fe)
of dust from $k$th type of stars ($k$ = I, II, and AGB for SNe Ia, SNe II, and
AGB stars, respectively) based on the methods described in B13a that is similar to
those adopted in Dwek (1998, D98).
We consider
that the key parameter in dust accretion is the dust accretion timescale ($\tau_{\rm a}$).
In the present study, this parameter can vary between different gas particles
and is thus represented by $\tau_{\rm a, \it i}$ for $i$th gas particle.
The mass of $j$th component
($j$=C, O, Mg, Si, S, Ca, and Fe) of dust for $i$th gas particle
at time $t$ ($d_{i,j}(t)$) can increase owing  to dust accretion processes.
The mass increase
is described as
\begin{equation}
\Delta d_{i,j}^{\rm acc}(t)=\Delta t_i (1-f_{\rm dust,\it i, j})
d_{i,j}(t) /\tau_{\rm a, \it i},
\end{equation}
where $\Delta t_i$ is the individual time step width for the $i$th gas particle
and $f_{\rm dust, \it i, j}$ is the fraction of the $j$th chemical element that
is locked up in the dust. Owing to this dust growth, the mass of $j$th chemical
component that is {\it not} locked up in the dust ($z_{i,j}(t)$)
can decrease, which is simply given as
\begin{equation}
\Delta z_{i,j}^{\rm acc}(t)=- \Delta t_i (1-f_{\rm dust,\it i, j})
d_{i,j}(t) /\tau_{\rm a, \it i}
\end{equation}
As is clear in these equations, the total  mass of $j$th component in $i$th gas
particle ($m_{i,j}(t)$) is $z_{i,j}(t)+d_{i,j}(t)$.

Dust grains can be destroyed though supernova blast waves
in the ISM of galaxies (e.g., McKee 1989)
and the destruction process is parameterized by the destruction time scale
($\tau_{\rm d}$) in previous one-zone models (e.g., Lisenfeld \& Ferrara 1998;
Hirashita 1999).  Following the previous models,
the decrease  of the mass of $j$th component
of dust for $i$th gas particle
at time $t$ due to dust destruction process
is as follows
\begin{equation}
\Delta d_{i,j}^{\rm dest}(t)= - \Delta t_i
d_{i,j}(t) /\tau_{\rm d, \it i},
\end{equation}
where $\tau_{\rm d, \it i}$ is the dust destruction timescale for $i$th particle.
The dust destroyed by supernova explosions can be returned back to the ISM,
and therefore the  mass
of $j$th chemical
component that is not locked up in the dust
increases  as follows:
\begin{equation}
\Delta z_{i,j}^{\rm dest}(t)= \Delta t_i
d_{i,j}(t) /\tau_{\rm d, \it i}
\end{equation}

Thus the equation for the time evolution of $j$th component of metals
for $i$th gas particle  are given as
\begin{equation}
z_{i,j}(t+\Delta t_i)=z_{i,j}(t)+\Delta z_{i,j}^{\rm ej}(t)+\Delta z_{i,j}^{\rm acc}(t)
+\Delta z_{i,j}^{\rm dest}(t)
\end{equation}
Likewise, the equation for dust evolution is given as
\begin{equation}
d_{i,j}(t+\Delta t_i)=d_{i,j}(t)+\Delta d_{i,j}^{\rm acc}(t)
+\Delta d_{i,j}^{\rm dest}(t)
\end{equation}
Dust is locked up in stars as metals are done so, when gas particles are converted into
new stars.

\subsubsection{Variable dust accretion models}

We investigate the 
variable dust accretion ('VDA')
model in which $\tau_a$ is different between different particles
with different gaseous properties and changes with time according to the changes
of gaseous properties.
We need to introduce a few additional parameters in VDA in order to
describe the possible dependences of $\tau_{\rm a}$  of gas particles
on the gas densities,
temperature, and chemical abundances.
Our previous simulations with VDA (B14)
clearly show the importance of dust accretion and destruction in the evolution
of $D$ and $f_{\rm H_2}$. 
The details of the VDA model is given in (B14), which discusses the comparison
between the VDA and the constant dust accretion model in which $\tau_{\rm a}$ is
constant for all particles throughout simulations.

We adopt the following dependence of $\tau_{\rm a, \it i}$ on the mass density
and temperature of a gas particle in the VDA:
\begin{equation}
\tau_{\rm a, \it i} = \tau_{\rm 0}  (\frac{ \rho_{\rm g,0} }{ \rho_{\rm g, \it i} })
(\frac{ T_{\rm g,0} } { T_{\rm g, \it i} })^{0.5},
\end{equation}
where $\rho_{\rm g, \it i}$ and $T_{\rm g, \it i}$ are the gas density and temperature
of a $i$-th  gas particle, respectively,
$\rho_{\rm a,0}$ (typical ISM density at the solar neighborhood)
and $T_{\rm g,0}$ (temperature of cold gas) are set to be 1 atom cm$^{-3}$ and 20K,
respectively, and $\tau_{\rm 0}$ is a reference dust accretion timescale
at $\rho_{\rm g,0}$ and $T_{\rm g, 0}$.
The dust destruction timescale,  $\tau_{\rm d, \it i}$, for each gas particle is 
described  as follows:
\begin{equation}
\tau_{\rm d, \it i} = \beta_{\rm d} \tau_{\rm a, \it i},
\end{equation} 
where $\beta$ controls the timescale ratio of dust destruction to dust growth. In our
previous studies (B13a, B14), $\beta_{\rm d}=2$ is demonstrated to reproduce the
observed dust properties of galaxies. We therefore consider that $\beta_{\rm d}$ should
be adopted a reference value in the present dust model.

In  the present study,
the dust-to-metal-ratio of a galaxy 
is defined as follows:
\begin{equation}
f_{\rm dust}=\frac{ M_{\rm dust} } { M_{\rm Z} },
\end{equation}
where $M_{\rm dust}$ and $M_{\rm Z}$ are the total amount  of dust and metals in the galaxy,
respectively. Although the initial value of $f_{\rm dust}$ could be different in different
galaxies, we mainly adopt the standard value of 0.4 (corresponding to the value of the 
solar neighborhood) and thereby investigate the evolution of dust and ${\rm H_2}$. We briefly
discuss how the present results depend on the initial $f_{\rm dust}$ by changing $f_{\rm dust}$
from 0.1 to 0.4.

\begin{figure*}
\psfig{file=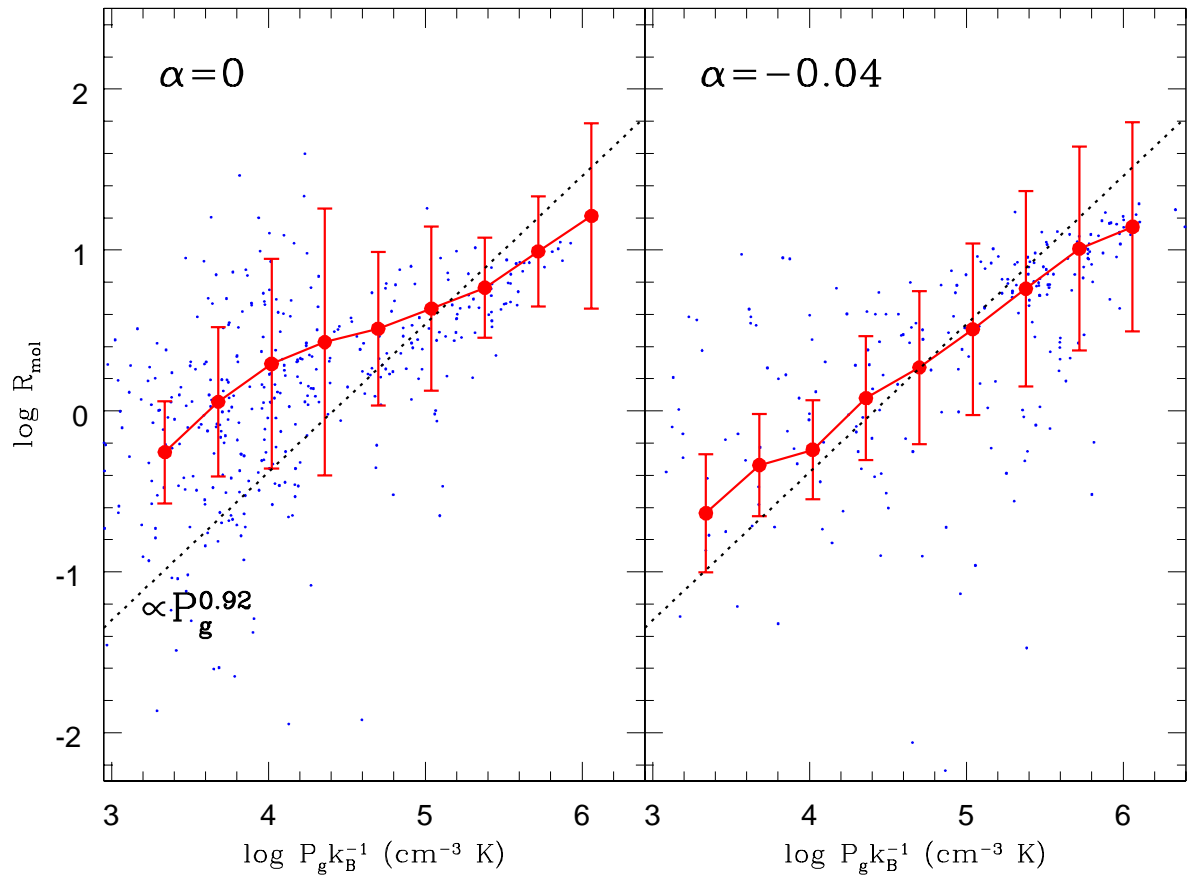,width=16.0cm}
\caption{
The plots of gas particles on the $R_{\rm mol}-P_{\rm g}$ plane in M1, where $R_{\rm mol}$
is the mass-ratio of ${\rm H_2}$ to H~{\sc i} and $P_{\rm g}$ is gaseous pressure.
Here $P_{\rm g} {\rm k}_{\rm B}^{-1}$ rather than $P_{\rm g}$ is  plotted for each gas particle so that 
the simulated correlation can be compared with the observed one (black dotted line) by
Blitz et al. (2007). The red big circles indicate the mean $R_{\rm mol}$ for each $P_{\rm g}$ bin
and the error bar  shows the dispersion in $R_{\rm mol}$. The models with $\alpha=0$ 
(flat metallicity gradient) and
$-0.04$ (steeper one) are shown in the left and right panels, respectively.
}
\label{Figure. 6}
\end{figure*}

\begin{figure*}
\psfig{file=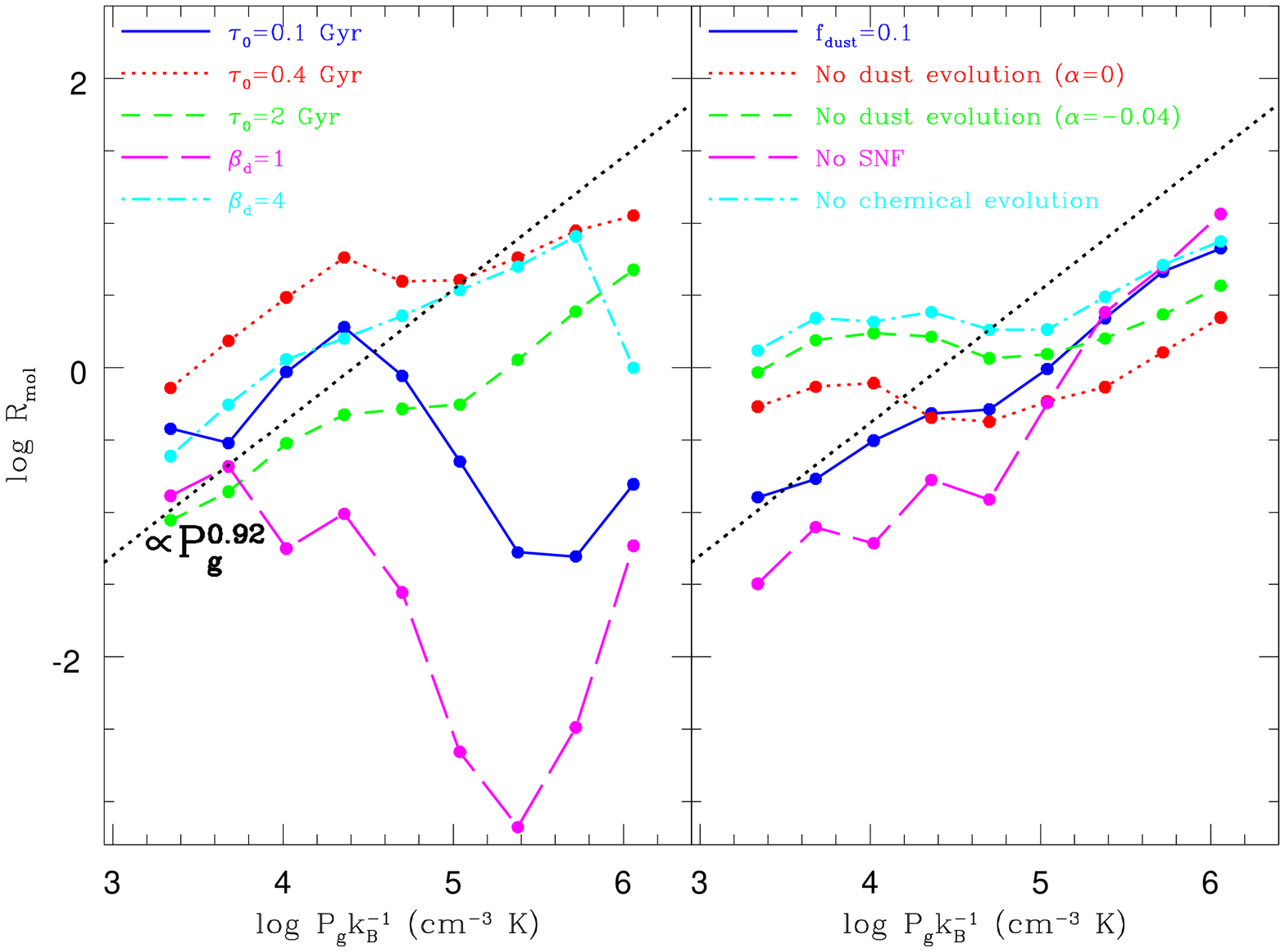,width=16.0cm}
\caption{
The same as as Fig. 5 but for different ten MW-type disk models (M1).
The left panel shows the models 
with  $\tau_0=0.1$ Gyr (blue solid), $\tau_0=0.4$ Gyr (red dotted),
$\tau_0=2$ Gyr (green short-dashed), $\beta_{\rm d}=1$ (magenta long-dashed), 
and $\beta_{\rm dust}=4$ (cyan dot-dashed).
The right panel shows the models 
with  $f_{\rm dust}=0.1$ Gyr (blue solid), no dust evolution and $\alpha=0$ (red dotted),
no dust evolution with $\alpha=-0.04$  (green short-dashed), no SN feedback effects (magenta long-dashed), 
and no chemical evolution  (cyan dot-dashed).
}
\label{Figure. 7}
\end{figure*}

\subsection{${\rm H_2}$ formation and dissociation}

The model for ${\rm H_2}$ formation and dissociation in the present study
is exactly the same as those used in B13a: ${\rm H_2}$ formation
on dust grains and ${\rm H}_2$ dissociation by FUV radiation
are both self-consistently included in chemodynamical simulations.
The temperature ($T_{\rm g}$),
hydrogen density ($\rho_{\rm H}$),  dust-to-gas ratio ($D$)
of a gas particle and the strength of the
FUV radiation field ($\chi$) around the gas particle
are calculated at each time step so that the fraction of molecular
hydrogen ($f_{\rm H_2}$) for the gas particle can be derived based on
the ${\rm H_2}$ formation/destruction equilibrium conditions.
Thus the ${\rm H_2}$ fraction for $i$-th gas  particle ($f_{\rm H_2, \it i}$)
is given as;
\begin{equation}
f_{\rm H_2, \it i}=F(T_{\rm g, \it i}, \rho_{\rm H, \it i}, D_i,  \chi_i),
\end{equation}
where $F$ means a function for $f_{\rm H_2, \it i}$ determination.

Since the detail of the derivation methods of $\chi_i$ and $f_{\rm H_2, \it i}$
(thus $F$)  are given
in B13a and B13b, we here briefly describe the methods.
The SEDs of stellar particles around each $i$-th gas particles 
(thus ISRF) are first 
estimated from ages and metallicities of the stars by using stellar population
synthesis codes for a given IMF (e.g., Bruzual \& Charlot 2003).
Then the strength of the FUV-part of the ISRF
is estimated from the SEDs so that $\chi_i$ can be derived for the $i$-th gas particle.
Based on $\chi_i$, $D_i$, and  $\rho_{\rm H, \it i}$ of the gas particle,
we can derive $f_{\rm H_2, \it i}$ (See Fig. 1 in B13a).
Thus each gas particle has $f_{\rm H_2, \it i}$, metallicity ([Fe/H]),
and gas density, all of which are used for estimating the IMF slopes
for the particle (when it is converted into a new star).

The ages of stars ($t_{\rm s}$) are assumed to be 5 Gyr and 2 Gyr for the models with $z=0$ and 
$z=2$, respectively. The results of the present models are influenced by the choice of 
$t_{\rm s}$ only for the very early-phase of disk evolution ($T<0.1$ Gyr), if $t_{\rm s}$
is rather short ($\sim 0.1$ Gyr). The influences of $t_{\rm s}$ is very limited, because
the photodissociation of ${\rm H_2}$ by stars can be very efficient when gas is surrounded by
very young stars and thus irradiated by the strong UV-radiation fields. We therefore discuss
only the results of the models with $t_{\rm s}$ adopted above.

A number of previous theoretical models of ${\rm H_2}$ formation (e.g., Hidaka \& Sofue 2002)
adopted the phase transition theory proposed by Elmegreen (1993), in which H~{\sc i}-to-${\rm H_2}$
transition in ISM  is determined basically by gas pressure and radiation field. Elmegreen (1993)
already showed that the observed ${\rm H_2}$ mass function ($N \propto  M^{-1.5}$) can be understood by
his theory and also suggested that spiral density waves can convert H~{\sc i} into ${\rm H_2}$.
We do not adopt his phase transition theory but instead try to reproduce the observed
${\rm H_2}$ properties using a different model based on ${\rm H_2}$ formation on dust grains.

\subsection{Gravitational dynamics and hydrodynamics}

One of key ingredients of the code is that
the gravitational softening length ($\epsilon$) is chosen for each
component in a galaxy (i.e.,
multiple gravitational softening lengths).
Thus the gravitational softening length ($\epsilon$)
is different between dark matter ($\epsilon_{\rm dm}$)
and gas ($\epsilon_{\rm g}$) and $\epsilon_{\rm dm}$ is determined by the initial
mean separation of dark matter particles.
The gravitational softening length for stars ($\epsilon_{\rm s}$) is set to be the same as that for gas.
Initial $\epsilon_{\rm g}$ is set to be significantly smaller than 
$\epsilon_{\rm dm}$ owing to rather  high number-density of gas particles.
The softening length for new stars formed from gas is set to be the same 
as $\epsilon_{\rm g}$.
Furthermore,  when two different components interact gravitationally,
the mean softening length for the two components
is applied for the gravitational calculation.
For example, $\epsilon = ({\epsilon}_{\rm dm}+{\epsilon}_{\rm g})/2$
is used for gravitational interaction between
The values of $\epsilon_{\rm dm}$ and $\epsilon_{\rm g}$ ($=\epsilon_{\rm s}$)
are 2.1 kpc and 0.2 kpc, respectively, for the fiducial MW-type disk model.

We consider that the ISM in galaxies can be modeled as an ideal gas with
the ratio of specific heats ($\gamma$) being 5/3.
The gaseous temperature ($T_{\rm g}$) is set to be $10^4$ K initially in all models.
The basic methods to implement SPH in the present study are essentially
the same as those proposed by Hernquist \& Katz (1989).
We adopt the predictor-corrector algorithm (that is accurate to second order
in time and space) in order to integrate the equations
describing  the time  evolution of a system.
Each particle is allocated an individual time step width ($\Delta t$) that is determined
by physical properties of the particle.
 The maximum time step width ($\Delta t_{\rm max}$)
is $0.01$ in simulation units, which means that  $\Delta t_{\rm max}=1.41 \times 10^6$ yr
in the present study. Although a gas particle is allowed to have a minimum time step
width of $1.41 \times 10^4$ yr in the adopted individual time step scheme,
no particle actually has such a short time step width ($\sim 10^4$ yr)  owing to conversion
from gas to star in high-density gas regions.
The radiative cooling processes
are properly included  by using the cooling curve by
Rosen \& Bregman (1995) for  $100 \le T < 10^4$K
and the MAPPING III code
for $T \ge 10^4$K
(Sutherland \& Dopita 1993).

\subsection{Tidal interaction model}

In order to investigate how external tidal perturbation can influence
the physical properties of ${\rm H_2}$ in disk galaxies,
we investigate tidal interaction models in which two disk galaxies
strongly interact with each other.
One of the two galaxies (`primary galaxy')
is represented by the disk galaxy model
described above whereas the interacting companion galaxy is represented
by a point-mass particle. Although
the mass-ratio of the companion to the primary can be  a free parameter represented
by $m_2$,
we present the results only for the models with $m_2=1$ in which the influences
of tidal interaction on ${\rm H_2}$ properties can be clearly seen.

In all of the simulations of tidal interaction, the orbit of the two disks is set to be
initially in the $xy$ plane and the distance between
the center of mass of the two disks
is set to b $10R_{\rm d}$ (corresponding to 175 kpc for the MW-type disk model) for most models.
The pericenter
distance, represented by $r_{\rm p}$, is set to be $3R_{\rm s}$.
The eccentricity
is  set to be 1.0 for all models of galaxy interaction, 
meaning that the encounter of galaxy interaction is parabolic.
The spin of the primary  galaxy in an interacting pair
is specified by two angle $\theta$ and
$\phi$ (in units of degrees), where
$\theta$ is the angle between the $z$ axis and the vector of
the angular momentum of a disk and
$\phi$ is the azimuthal angle measured from $x$ axis to
the projection of the angular momentum vector of a disk on
to $xy$ plane.

In the present study, we present the results of the following
four tidal interaction models: (i) prograde (`PR') 
model with $\theta$ = 0, $\phi$ = $0$, and $r_{\rm p}=3R_{\rm s}$,
(ii) retrograde (`RE') model with
$\theta$ = 180,  $\phi$ = 0, and $r_{\rm p}=3R_{\rm s}$
(iii) highly inclined model
with $\theta$ = 30, $\phi$ = $60$, and $r_{\rm p}=3R_{\rm s}$,
and  (iv) distant interaction model
with $\theta$ = 0, $\phi$ = $0$, and $r_{\rm p}=4R_{\rm s}$.

\subsection{A parameter study to solve key questions}

We investigate numerous models to try to answer the following key questions
related to the origin of ${\rm H_2}$ in disk galaxies: (i) how can ${\rm H_2}$ gas
distribute within disks ?, (ii) what determines the molecular gas (${\rm H_2}$) fractions
in disk galaxies ?, (iii) what are the roles of galactic bulges in controlling ${\rm H_2}$
properties of disk galaxies ?, (iv) can dust evolution influence the evolution
of H~{\sc i} and ${\rm H_2}$ in galaxies ?,  (v) does galaxy interaction enhance
${\rm H_2}$ contents of galaxies ?, (vi) is there any threshold gas density (or
threshold galaxy mass) beyond which ${\rm H_2}$ formation is possible ?,
and (vii) can high-$z$ disk galaxies can contain larger fractions of ${\rm H_2}$ within
disks ? It should be noted here that these are selected, because the present numerical code
allows us to investigate these: other questions such as the influences of AGN and dust size evolution
on ${\rm H_2}$ evolution can not be addressed in the present study.

In order to address these questions, we mainly describe the results of the selected 34 models 
in the present study. We indeed investigated more than 34 models, in particular,
for low-mass disk models in order to understand whether there can be
a threshold halo mass beyond which ${\rm H_2}$ formation is possible. 
However, we discuss only these results, because these
representative  models can more clearly show how the results
depend on model parameters.
The values of parameters adopted in these models are given in Table 2. 
In order to discuss the importance of dust evolution in ${\rm H_2}$ evolution
of galaxies, we investigate the models with different basic dust parameters.
The parameter values for dust models and chemical evolution are summarized in Table 3.
We first describe the results of the `fiducial model' (M1) that corresponds to
a disk galaxy similar to the Milky Way (thus referred to as MW-type disk model).
Then we discuss how the key model parameters controls the ${\rm H_2}$ properties
of disk galaxies in detail.

\subsection{A method to estimate ${\rm H_2}$ surface density distribution (H2SDD) }

The present chemodynamical models do not  have enough spatial resolution to resolve the sub-pc-scale
central cores of GMCs with ${\rm H_2}$. We therefore can not discuss whether and where
self-gravitating molecular clouds are formed in galactic gas disks.
Instead of investigating the molecular cloud mass function (MCMF),
we try to derive the ${\rm H_2}$ surface density distribution (H2SDD),
which would be useful in discussing the origin of MCMFs of galaxies.
We  estimate the H2SDD
in a disk galaxy as follows.
We first divide the simulated stellar disk region  ($R_{\rm s}=17.5$ kpc for the MW model) of a disk galaxy
into $200 \times 200$ or $400 \times 400$ small meshes and thereby estimate the total mass
of ${\rm H_2}$ ($m_{\rm H_2, mesh}$). 

We count the number of meshes ($N_{\rm H_2}$) with $m_{\rm H_2, mesh}$
within a given mass range so that we can estimate the H2SDD.
The mesh size can be as small as $\sim 90$pc for the MW-type disk models, which can be enough
to discuss the mass of each individual ${\rm H_2}$ gas clouds  in a local region of a disk galaxy.
Although this method is not exactly the same as that used in the observational estimation of  MCMF,
we consider that the derived H2SDD could be  quite useful in 
discussing the origin of 
the observed MCMF in a qualitative manner (e.g., discussing the slope of the MCMF).
We mainly check whether the simulated slope of H2SDD can be similar to the observed slope of the
Galactic MCMF.

\begin{figure*}
\psfig{file=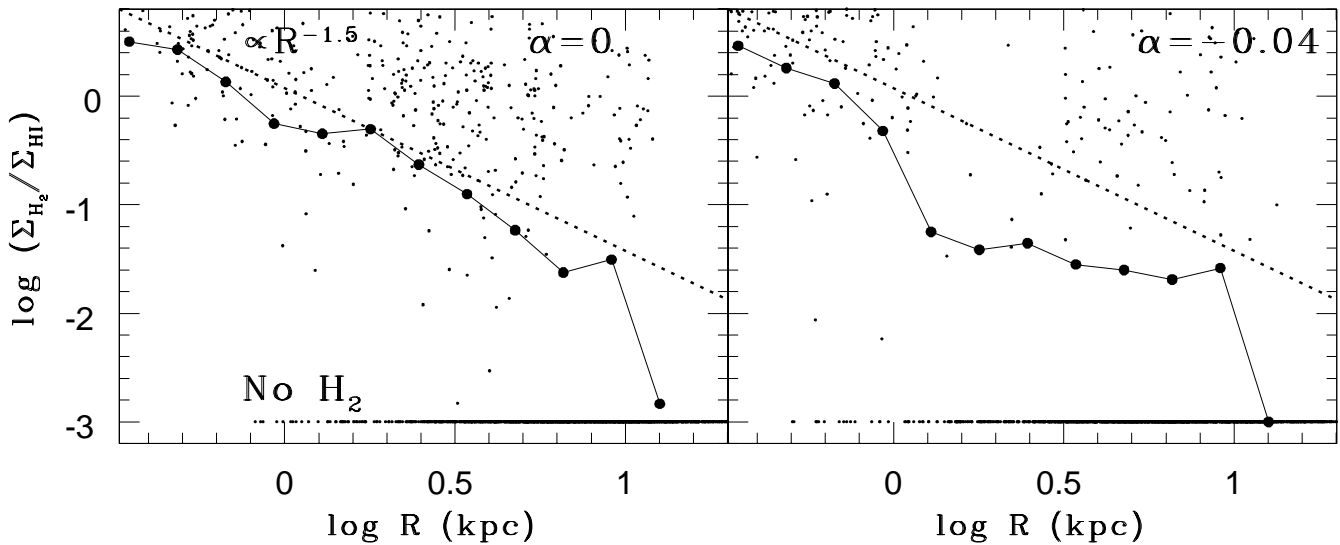,width=18.0cm}
\caption{
The projected radial distribution of the surface-density-ratios of ${\rm H_2}$ to H~{\sc i}
at $T=1.1$ Gyr in the fiducial MW-type disk model (M1) with $\alpha=0$ (left) and $\alpha=-0.04$ (right).
The observed correlation ($\Sigma_{\rm H_2}/\Sigma_{\rm HI} \propto R^{-1.5}$) by Wong \& Blitz (2002)
is shown by a dotted line for comparison. The small dots represent the locations of the selected
gas particles and the particles with no ${\rm H_2}$ (i.e., $f_{\rm H_2}=0$) are shown at
$\log(\Sigma_{\rm H_2}/\Sigma_{\rm HI})=-3$ for convenience.
}
\label{Figure. 8}
\end{figure*}

\section{Results}

\subsection{Fiducial MW model}

\subsubsection{Spatial distributions  of hydrogen gas
and ${\rm H_2}$ surface density distribution}

Figs. 1 and 2 describe how the spatial distributions of stars, H~{\sc i}, ${\rm H_2}$,
and dust evolve with time during the dynamical evolution of the gas disk
with spiral arms and a central bar  in the fiducial MW-type disk model (M1).
Numerous spiral arms can be developed at $T=0.6$ Gyr owing to gravitational instability,
and they can influence the conversion from H~{\sc i} to ${\rm H_2}$ in the gas disk after their formation.
The locations  of high-density ${\rm H_2}$ regions at $T=0.6$, 0.8, and 1.1 Gyr appear
to be roughly coincident with the locations of H~{\sc i} spiral arms
of the gas disk.
However, ${\rm H_2}$ shows a very clumpy distribution along spirals arms,
which is in a striking contrast with the relatively smooth distribution of H~{\sc i} along spiral arms.
This implies that only the high-density parts of the gaseous spiral arms can be the possible
formation sites of ${\rm H_2}$ in galactic disks.
Numerous ${\rm H_2}$ clumps can be active star-forming regions in the present SF model
based on ${\rm H_2}$ gas densities.
Although the projected distribution of interstellar dust follows the spiral-like distribution
of H~{\sc i} gas, it does not show clump-like structures as ${\rm H_2}$.
Interstellar dust can become less clumpy, because it can be efficiently destroyed by star-formation (i.e., SNe)
which can occur in  high-density gas clumps.

The physical roles of spiral arms in the formation of  GMCs have been already discussed in
a number of recent theoretical works based on high-resolution numerical simulations of gas disk
evolution in disk galaxies (e.g., Dobbs et al. 2011; Wada et al. 2011; Halle \& Combes 2013).
Although these previous works did not self-consistently model the ${\rm H_2}$ formation on dust
grains and the time evolution of dust,  they have already shown projected ${\rm H_2}$ 
distributions in galactic disks (e.g., Fig. 14 in Halle \& Combes 2013). The derived distributions
of ${\rm H_2}$ along spiral arms in the present  MW-type disk model
are similar to  those reported in Halle \& Combes (2013).
The clumpy ${\rm H_2}$ distributions derived in the present study
are not so clearly seen in previous works, which could reflect
the differences in the models of SF and SN feedback effects between the present and other works.

\begin{figure*}
\psfig{file=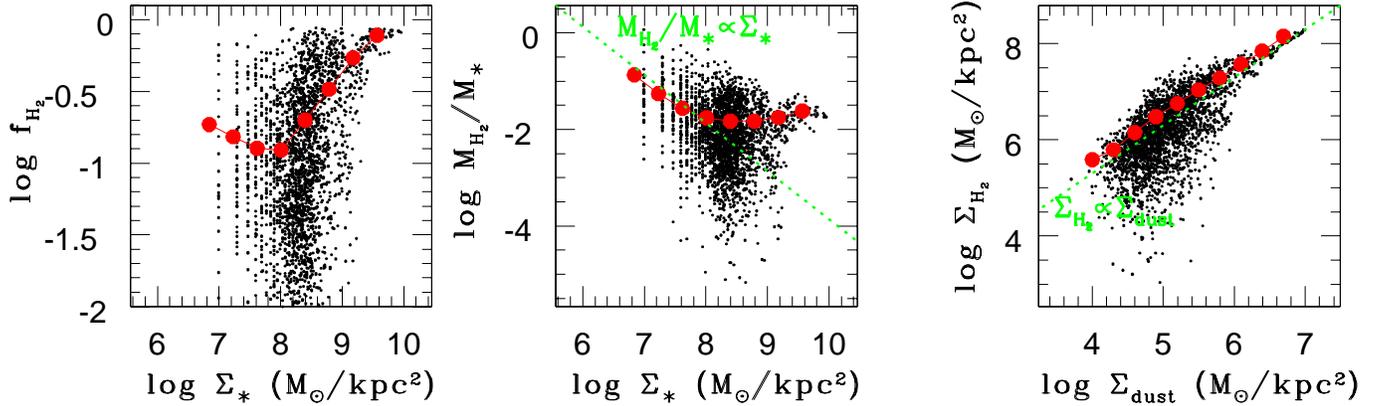,width=18.0cm}
\caption{
The simulated three ${\rm H_2}$-scaling relations at $T=1.1$ Gyr in the fiducial MW-type disk model (M1).
The left, middle, and right panels show correlations between $f_{\rm H_2}$ and $\Sigma_{\ast}$,
$M_{\rm H_2}/M_{\ast}$ and $\Sigma_{\ast}$, and $\Sigma_{\rm H_2}$ and $\Sigma_{\rm dust}$, 
respectively. Small black dots indicate the physical properties (e.g., $f_{\rm H_2}$) of local
regions (represented by gas particles) whereas big red dots indicate the average values at each bin.
For comparison, the linear correlations of $M_{\rm H_2}/M_{\ast} \propto \Sigma_{\ast}$ and
$\Sigma_{\rm H_2} \propto \Sigma_{\rm dust}$ are shown by green dotted lines in the middle and right
panels, respectively.
}
\label{Figure. 9}
\end{figure*}

Fig. 3 shows the spatial distributions of H~{\sc i}
and ${\rm H_2}$ 
and the mass function of ${\rm H_2}$ gas clouds 
(referred to as  molecular cloud mass function; MCMF)
in the gas disk at $T=1.1$ Gyr 
in the fiducial MW-type disk
galaxy model with $f_{\rm g}=0.09$.
The simulated  ${\rm H_2}$  distribution
is more centrally concentrated than H~{\sc i}
gas within the disk owing to the more efficient
formation of ${\rm H_2}$ on dust grains in the inner part of the disk
where both gas density and $D$ can be higher during the disk evolution.
The spatial distribution of ${\rm H_2}$ gas shows  small-scale clumpy
structures within gaseous spiral arms where gas density can become
rather large. Clearly, there is an outer truncation 
($R \sim 13$ kpc) beyond which
no/little ${\rm H_2}$ gas can be found in this model.
The projected radial density profile of ${\rm H_2}$ 
is steeper than that of H~{\sc i} in this model.

The central  part of the disk is dominated by ${\rm H_2}$ gas 
and most of the gas particles there have rather high $ f_{\rm  H_2}$ 
($>0.5$), which implies that molecular hydrogen can be systematically
more massive.
The simulated  H2SDD
has a slope of $\sim -1.5$ for $\log \Sigma_{\rm H_2}<2.4$ ${\rm M}_{\odot}$ pc$^{-2}$,
which means that the simulated slope is 
similar to the observed slope of MCMF  that is approximated as
$N_{\rm mc}  \propto m_{\rm mc}^{-1.5}$ for the MW  and M31
(e.g., Fig. 4 in Fukui \& Kawamura 2010).
This  implies that the present model of
${\rm H_2}$ formation on dust grains is realistic and reasonable for ${\rm H_2}$ formation
in disk galaxies.
Although the MCMF can be shifted toward higher masses if 
a coarser mesh size (${\rm n}_{\rm mesh}=200^2$) is used for the MCMF derivation,
the slope of the MCMF does not change significantly.

This MW-type fiducial model can  develop a central stellar bar owing
to  bar instability and the stellar bar can influence the formation
process of ${\rm H_2}$ significantly. 
The formation process of the stellar  bar in the central 2 kpc
and its influence on gas for $T \le 1.1$ Gyr are  given in Appendix A.
Fig. 4 shows that the gas disk has  a very strong concentration of ${\rm H_2}$
in the central 200 pc at $T=2.3$ Gyr. This ${\rm H_2}$ concentration 
corresponds to the formation of nuclear ${\rm H_2}$ gas disk due to the
dynamical action of the central stellar bar on the surrounding gas.
Owing to the rather high concentration of ${\rm H_2}$ in the disk,
the SFR can be significantly increased, in particular, in the central region
of this barred disk galaxy. Furthermore, the ${\rm H_2}$ fraction
can be significantly increased (from $f_{\rm H_2} \sim 0.1$ to $\sim 0.3$).
Thus these results clearly demonstrate that stellar bars can play a vital role
in converting H~{\sc i} to ${\rm H_2}$ in disk galaxies.

Fig. 5 describes how the simulated H2SDDs depend on the model parameters for 
metallicity gradient and dust growth/destruction. The slopes of the H2SDDs
do not depend strongly
on the parameters, though some models (with steep metallicity gradient
and longer dust accretion/destruction timescales)  show the overproduction of high-density
${\rm H_2}$ regions  with $\log \Sigma_{\rm H_2}>2.4$ ${\rm M}_{\odot}$ pc$^{-2}$ within their disks.
The larger number of high-density ${\rm H_2}$ regions
is due largely to the stronger 
central concentration of ${\rm H_2}$ gas in the inner regions of the 
gas disks. As shown in  Fig. 4, a central region of a galaxy is dominated by high-density
${\rm H_2}$ gas so that  almost all of the  meshes in the central region can have high $\Sigma_{\rm H_2}$.
As a result of this,  the simulated H2SDDs can have a bump around $\log \Sigma_{\rm H_2} = 2.6-2.8$
${\rm M}_{\odot}$ pc$^{-2}$. 
It is a bit surprising
that the simulated  H2SDDs  are less sensitive to the adopted dust model.

\begin{figure*}
\psfig{file=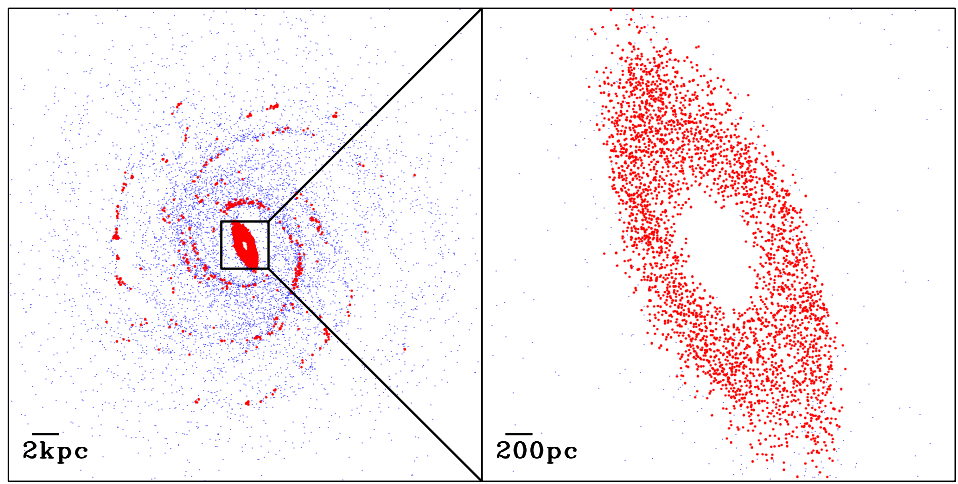,width=14.0cm}
\caption{
The distributions of gas with $f_{\rm H_2}<0.01$ (blue, H~{\sc i})
and $f_{\rm H_2} \ge 0.01$ (red, ${\rm H_2}$) projected onto the $x$-$y$ plane at $T=2.3$ Gyr in
the big bulge model with $f_{\rm b}=1$ (M8).
Clearly, an elongated ${\rm H_2}$-ring can be seen in this model.
}
\label{Figure. 10}
\end{figure*}
 
\begin{figure*}
\psfig{file=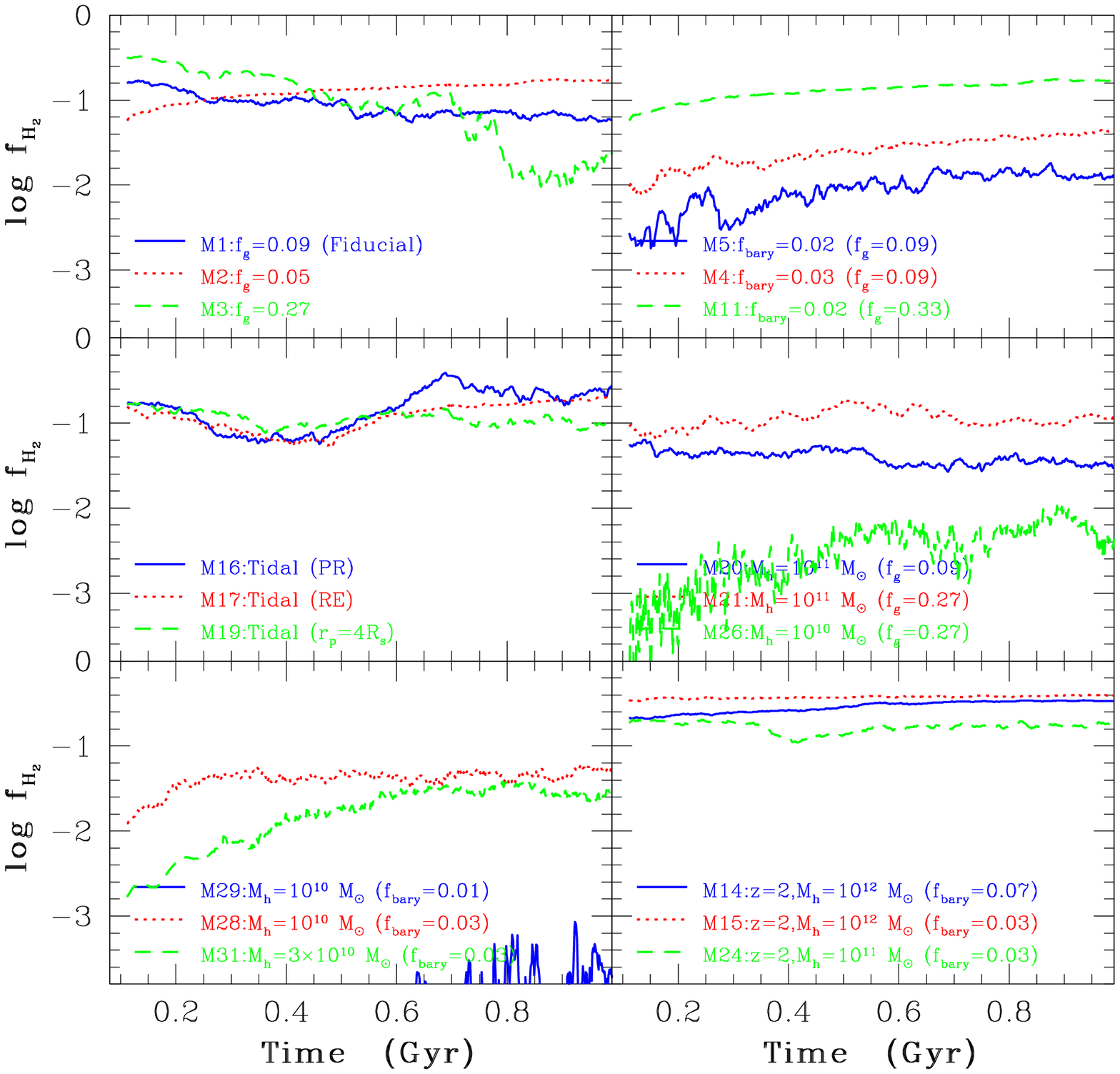,width=18.0cm}
\caption{
The time evolution of $f_{\rm H_2}$ in the representative 18 isolated
and interaction models with different $f_{\rm g}$,
$f_{\rm bary}$,  $M_{\rm h}$,
and $z$.  Each panel includes only three models in which only one or two model parameters
are different so that the roles of the parameter(s) in controlling the time evolution of $f_{\rm H_2}$
can be more clearly shown. For example, the top left panel describes how $f_{\rm g}$ 
can control $f_{\rm H_2}$ in disk galaxies.
}
\label{Figure. 11}
\end{figure*}

\subsubsection{The fraction of molecular hydrogen}

Fig. 6 describes the physical correlations between
molecular fraction ($R_{\rm mol}$), which is defined as
the mass ratio of ${\rm H_2}$ to H~{\sc i}, and $P_{\rm g} k_{\rm B}^{-1}$,
where $P_{\rm g}$ is the gas pressure of a particle and
$k_{\rm B}$ is the Boltzmann constant, in the gas disk of the fiducial
MW model with different initial metallicity gradients $\alpha=0$ 
and $-0.04$ dex kpc$^{-1}$ at $T=1$ Gyr. 
It should be stressed that this $R_{\rm mol}$ is not exactly the same as
$f_{\rm H_2}$ ($=M_{\rm H_2}/(M_{\rm HI}+M_{\rm H_2})$), which is often shown in 
figures of this paper.
The observed correlation of $R_{\rm mol} \propto P_{\rm g}^{0.92}$
(e.g., Blitz et al. 2007; Fukui \& Kawamura 2010) can be used as a key observational
constraint  for the present
${\rm H_2}$ formation model.
Although the dispersion
in $R_{\rm mol}$ at each bin appears to be large, 
the two models show a clear positive correlation between
$R_{\rm mol}$ and $P_{\rm g}$ (i.e.,  higher $R_{\rm mol}$ for
higher gaseous pressure). This result implies that the present model
for ${\rm H_2}$ formation, which does not explicitly assume a dependence
of ${\rm H_2}$ formation efficiency on gaseous pressure, does a good job
in predicting $R_{\rm mol}$ and its dependence on gaseous pressure.

Although the two models with different slopes of gaseous metallicity
gradients ($\alpha=0$ and $-0.04$) have the slopes
of the $R_{\rm mol}-P_{\rm g}$
that are similar to the observed one ($R_{\rm mol} \propto P_{\rm g}^{-0.92}$),
the model with a steep negative gradient of gaseous metallicity
($\alpha=-0.04$) can have the slope
more similar to the observed one.
The model with $\alpha=-0.04$ shows 
a lower ${\rm H_2}$ formation efficiency at 
$P_{\rm g} {\rm k}_{\rm B}^{-1} < 10^5$
(cm$^{-3}$ K) in comparison with the model with $\alpha=0$, mainly because
the model with a negative metallicity gradient has a significantly lower
dust-to-gas-ratio (i.e., lower metallicity) at the outer region
of the disk with lower  gaseous pressure so that
${\rm H_2}$ formation efficiency on dust grains can become lower.
The derived weaker dependence of the slope of the $R_{\rm mol}-P_{\rm g}$
relation on initial metallicity gradients
implies that the slopes can be quite similar between
different galaxies with different metallicity gradients.

Fig. 7 describes the simulated $R_{\rm mol}-P_{\rm g}$ relations in models
with different model parameters for dust growth and destruction.
These models have the parameter values being less realistic 
(e.g., rather long  dust accretion timescale) in comparison
with the models used in Fig. 6 so that we can discuss 
how the dust modeling is important for reproducing the observed
$R_{\rm mol}-P_{\rm g}$ relation.
It is clear from Fig. 7 that the simulated
$R_{\rm mol}-P_{\rm g}$ relation in the model with shorter dust growth timescale
($\tau_0=0.1$ Gyr) is less consistent with the observed one owing to
rather small $R_{\rm mol}$ at higher $P_{\rm g}$ ($>10^5 {\rm k}_{\rm B}$)
in the model. The model with a shorter dust destruction timescale
($\beta_{\rm d}=1$ and $\tau_0=0.2$ Gyr) can not reproduce
the observed $R_{\rm mol}-P_{\rm g}$ relation so well either.
These results imply that both dust accretion and destruction timescales
should be carefully chosen for reproducing the observed
$R_{\rm mol}-P_{\rm g}$ relation.

The models with longer dust accretion timescale ($\tau_0=0.4$ and 2 Gyr)
show  $R_{\rm mol}-P_{\rm g}$ similar to the observed one as long as
$\beta_{\rm d}=2$ is adopted.  However these models can not reproduce 
the observed $R_{\rm mol}-P_{\rm g}$ better than the models in Fig. 6, which
implies that not only $\beta_{\rm d}$ but also $\tau_0$ can be a key
parameter for reproducing the observed  $R_{\rm mol}-P_{\rm g}$.
The slope of the simulated  $R_{\rm mol}-P_{\rm g}$ in the
model with a longer dust destruction time scale ($\beta_{\rm d}=4$
and $\tau_0=0.2$ Gyr) can better  match the observed one except for
rather high gaseous pressure ($P_{\rm g} \sim 10^6 {\rm k}_{\rm B}$).  
The model with an initially lower dust-to-metal-ratio ($f_{\rm dust}=0.1$)
shows $R_{\rm mol}-P_{\rm g}$ relation similar to those in Fig. 6,
though $R_{\rm mol}$ at higher $P_{\rm g}$ is systematically smaller
than those in Fig. 6 owing to the lower ${\rm H_2}$ formation efficiency
caused by the  lower dust-to-gas ratio.

The models with no dust evolution show significantly shallower 
$R_{\rm mol}-P_{\rm g}$ relations, irrespective of the slopes of
the initial radial metallicity gradients ($\alpha=0$ and $-0.04$).
This result implies that the time evolution of dust abundances in 
disk galaxies needs to be included for the self-consistent reproduction
of the observed $R_{\rm mol}-P_{\rm g}$ relation.
The models without SN feedback (SNF)  and those without chemical evolution show 
significant deviation from the observed 
$R_{\rm mol}-P_{\rm g}$, which strongly suggests that 
dust growth and destruction processes caused by chemical enrichment
and SN explosions need to be carefully included in reproducing
the observed $R_{\rm mol}-P_{\rm g}$ relation in a quantitative manner.
Thus these results demonstrate that as long as the ${\rm H_2}$ formation
processes on dust grains are included in hydrodynamical simulations
of galaxy formation and evolution, 
the dust growth and destruction processes, which can be 
influenced by chemical enrichment and SN explosions,
should be carefully considered in discussing the origin
of the observed $R_{\rm mol}-P_{\rm g}$ relation.

\subsubsection{Radial gradients of molecular fraction}

The observed radial gradients of $R_{\rm mol}$ in disk galaxies
(e.g., Wong \& Blitz 2002)  can be an additional
constraint for any model for ${\rm H_2}$ formation in galaxy-scale
simulations. We here estimate $R_{\rm mol}$ at each radius in a simulated
disk galaxy by calculating the ratio of $\Sigma_{\rm H_2}$
(i.e., the projected surface mass density of ${\rm H_2}$)
to $\Sigma_{\rm HI}$ at each radius in Fig. 8.
Although individual gas particles corresponding to individual 
local gaseous regions in a disk galaxy
have vastly different $R_{\rm mol}$ (ranging from 0 to 10),
the slopes of the radial $R_{\rm mol}$ profiles
in the two models with different  $\alpha$ appear to be similar to the observed
one ($R_{\rm mol} \propto R^{-1.5}$).
The model with $\alpha=0$
can slightly better reproduce the observed slope than the model 
with $\alpha=-0.04$. Thus, the simulated slopes similar to the observed
$R_{\rm mol} \propto R^{-1.5}$ demonstrate that the present model
for ${\rm H_2}$ formation on dust grains is quite reasonable and realistic
in discussing the global (galaxy-scale) distributions of neutral and
molecular hydrogen in disk galaxies (as long as a reasonable set of
model parameters for dust is adopted).

\subsubsection{${\rm H_2}$ scaling relations}

A number of recent observational studies have found intriguing correlations
between physical properties of dust, gas,  and stars in galaxies (e.g., Corbelli et al 2012;
Cortese et al. 2012; Boselli et al. 2014).
It is
therefore essential for the present study to provide some comparisons
between these observations and the corresponding simulation results.
Fig. 9 shows that the ${\rm H_2}$ fractions ($f_{\rm H_2}$)
are more likely to be larger
for local regions with higher stellar surface densities ($\Sigma_{\ast}$),
in particular, for $\Sigma_{\ast} >10^8 {\rm M}_{\odot}$ kpc$^{-2}$.
There is a large dispersion in $f_{\rm H_2}$ for a given
$\Sigma_{\ast}$, which reflects the fact that star formation histories
and chemical enrichment processes (thus dust growth/destruction processes)
are quite different in different local regions. 
The local regions with higher $\Sigma_{\ast}$ can have higher gas densities
so that conversion from H~{\sc i} to ${\rm H_2}$ can occur more efficiently.
This is a physical reason for the simulated trend of $f_{\rm H_2}$ with
$\Sigma_{\ast}$ in Fig. 9.

There is a positive correlation between the
mass-ratio of ${\rm H_2}$ to stars and $\Sigma_{\ast}$, though
the correlation can not be simply described as a power-law profile owing
to the flatter slope in the simulated relation 
at $\Sigma_{\ast} >10^8 {\rm M}_{\odot}$ kpc$^{-2}$.
A large dispersion in $M_{\rm H_2}/M_{\ast}$ can be seen, which again
reflects the diverse histories of star formation and chemical enrichment
in different local regions. The simulated relation
between $\Sigma_{\rm H_2}$ and $\Sigma_{\rm dust}$ is relatively tight
in comparison with other two correlations and can be described roughly
as $\Sigma_{\rm H_2} \propto \Sigma_{\rm dust}$.
These simulated relations between physical properties of dust, gas,
and stars are at least qualitatively consistent with observational
results (e.g., Boselli et al. 2014).

\begin{figure*}
\psfig{file=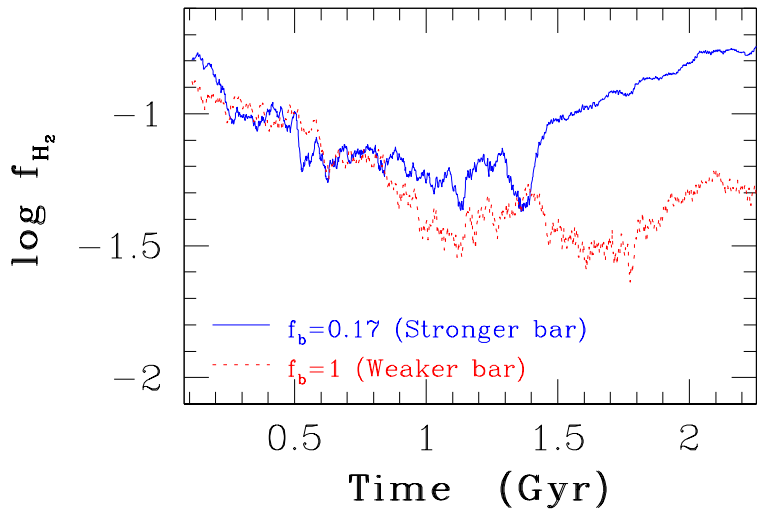,width=10.0cm}
\caption{
The long-term  evolution of $M_{\rm H_2}$ (upper) and $f_{\rm g}$ (lower)
in the two comparative models with $f_{\rm b}=0.167$ (M1, blue solid, smaller bulge)
and $f_{\rm b}=1$ (M8, red dotted, bigger bulges).
The disks with smaller bulges are more likely to develop stronger bars in their central
regions owing to global bar instability in the present study. The stronger bars can dynamically
act on gas disks so that a significant amount of gas can be transferred to the inner regions,
where high-density ${\rm H_2}$ regions can form. 
The later increase of $f_{\rm H_2}$ in the fiducial model with $f_{\rm b}=0.167$ is due
to the central gas concentration in the disk caused by dynamical action of the bar
(see also Fig. 4).
}
\label{Figure. 12}
\end{figure*}

\begin{figure*}
\psfig{file=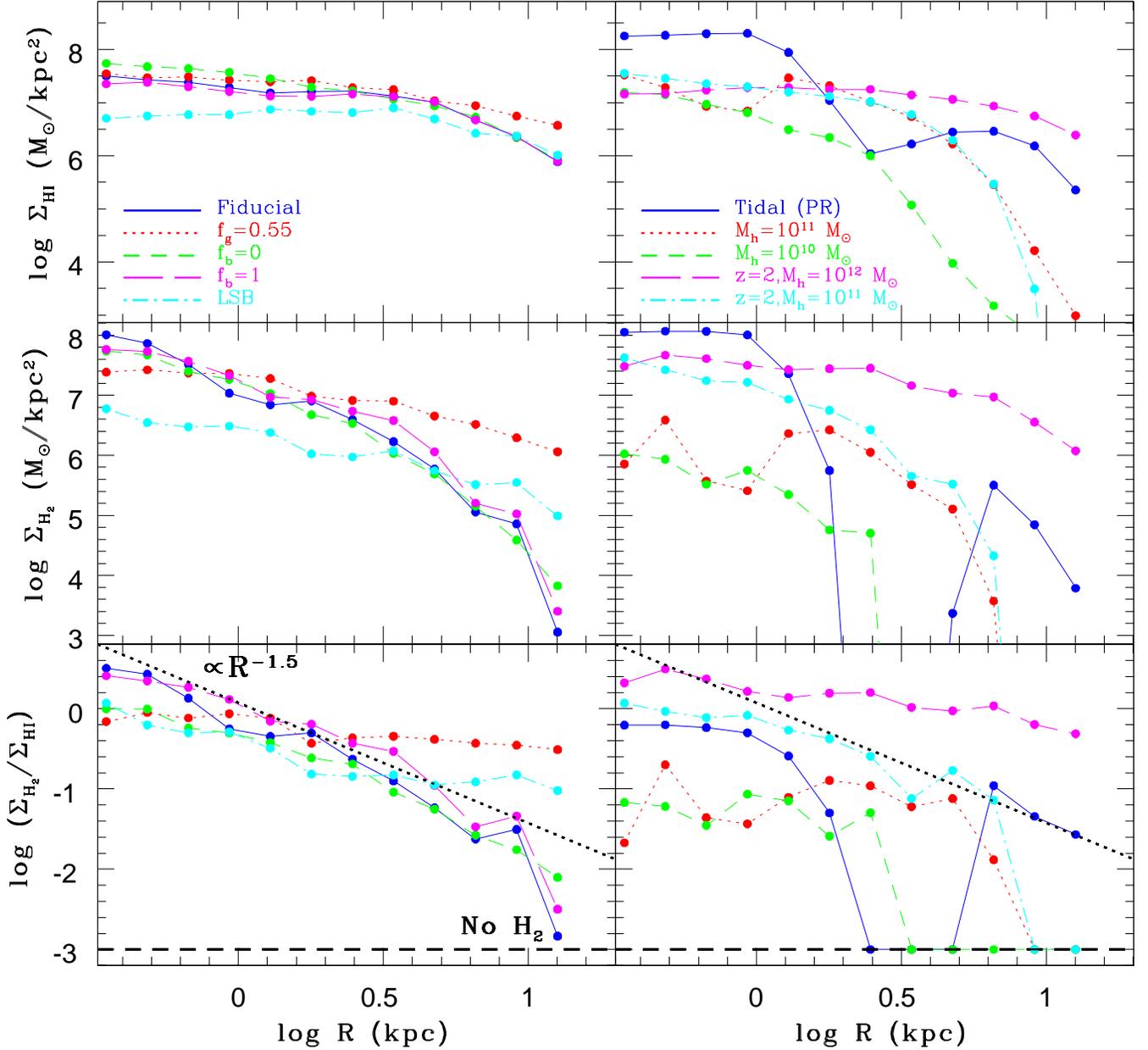,width=18.0cm}
\caption{
The projected radial density profiles of H~{\sc i} (top), ${\rm H_2}$ (middle),
and ${\rm H_2}$-to-H~{\sc i}-ratio for different ten models.
The left panel shows the MW-type disk models  ($M_{\rm h}=10^{12} {\rm M}_{\odot}$):
M1 (blue solid, fiducial), 
M6 (red dotted), 
M7 (green short-dashed), 
M8 (purple long-dashed), 
and M13 (cyan dot-dashed, LSB).
The right panel shows the five different models:
M16 (blue solid),  
M21 (red dotted),
M27 (green short-dashed), 
M15 (magenta long-dashed), 
and M24 (cyan dot-dashed).
}
\label{Figure. 13}
\end{figure*}

\begin{figure*}
\psfig{file=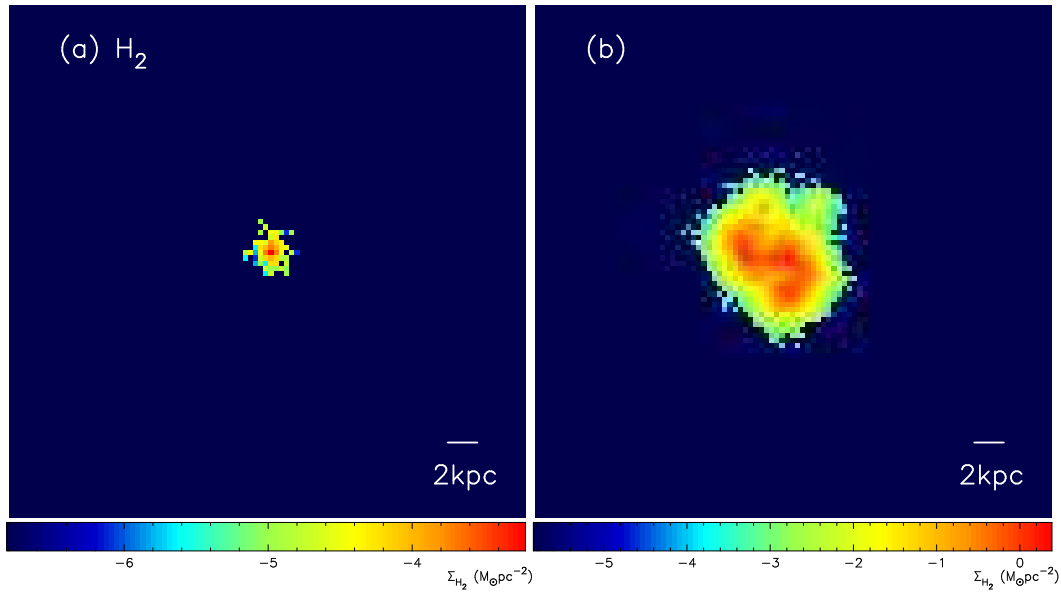,width=14.0cm}
\caption{
Th  projected 2D ${\rm H_2}$ density distributions ($\Sigma_{\rm H_2}$) 
for the low-mass dwarf disk models ($M_{\rm h}=10^{10} {\rm M}_{\odot}$), 
M25 with $f_{\rm g}=0.09$  ((a), left)
and M26 with $f_{\rm g}=0.27$  ((b), right). Clearly, $\Sigma_{\rm H_2}$ is much
lower in these low-mass models in comparison with the fiducial MW-type disk model.
}
\label{Figure. 14}
\end{figure*}

\begin{figure*}
\psfig{file=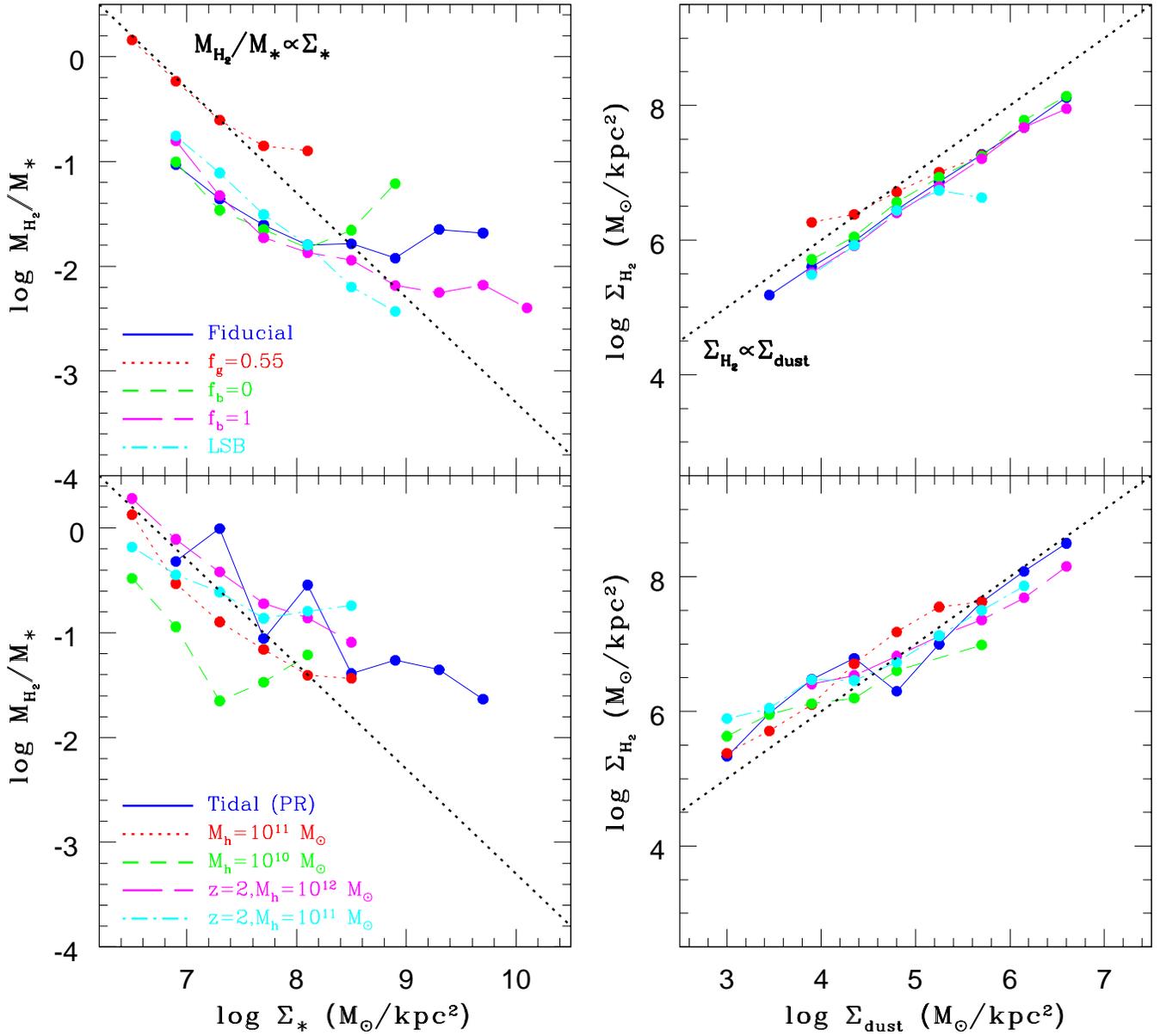,width=18.0cm}
\caption{
The simulated ${\rm H_2}$-scaling relations for the representative ten models,
which are the same as shown in Fig. 13. 
The left and right panels show correlations between 
$M_{\rm H_2}/M_{\ast}$ and $\Sigma_{\ast}$ and $\Sigma_{\rm H_2}$ and $\Sigma_{\rm dust}$, 
respectively.
For comparison, the linear correlations of $M_{\rm H_2}/M_{\ast} \propto \Sigma_{\ast}$ and
$\Sigma_{\rm H_2} \propto \Sigma_{\rm dust}$ are shown by black  dotted lines in the left and right
panels, respectively.
}
\label{Figure. 15}
\end{figure*}

\begin{figure*}
\psfig{file=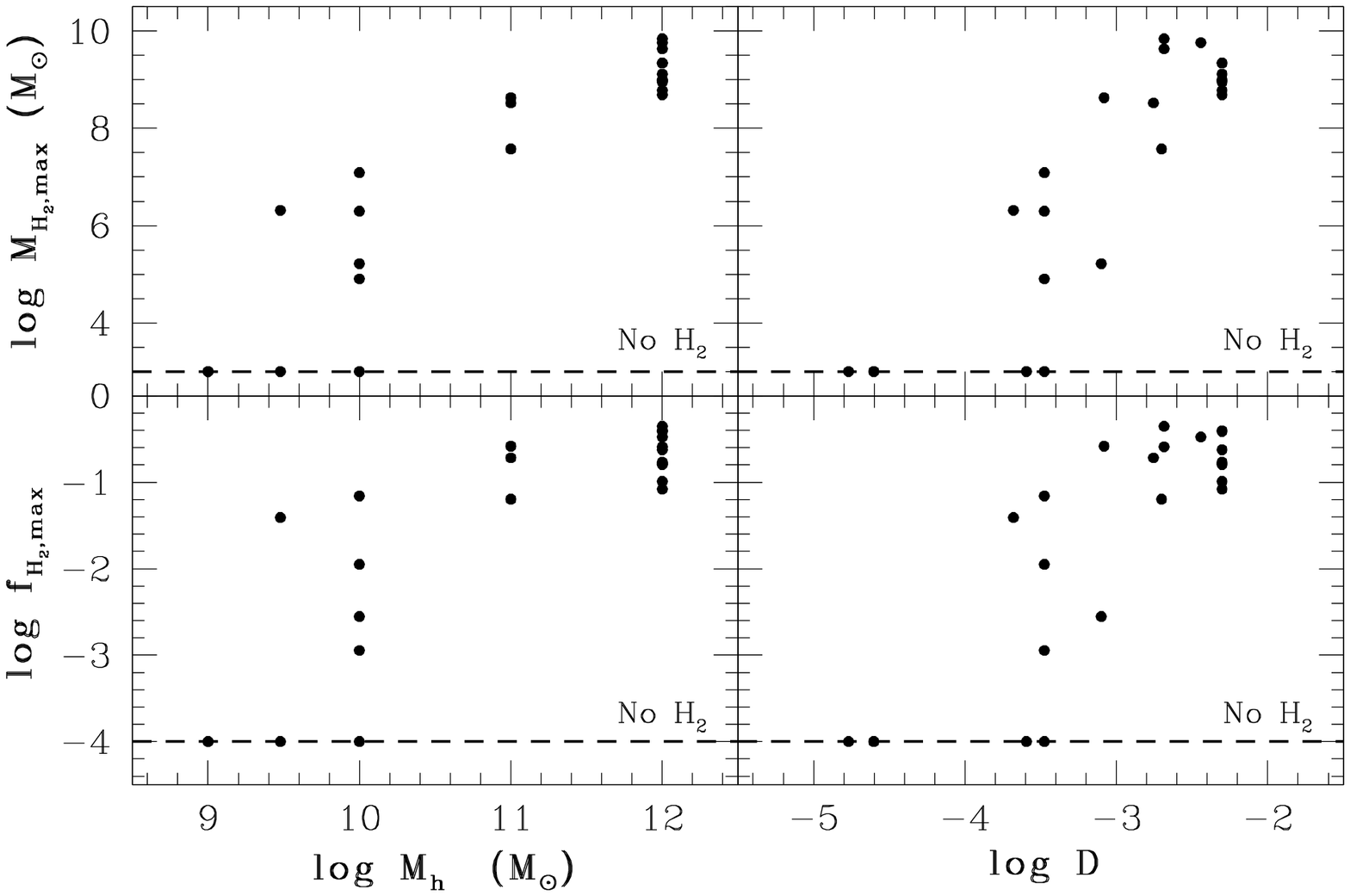,width=16.0cm}
\caption{
The dependences of the maximum $M_{\rm H_2}$ ($M_{\rm H_2, max}$, upper) and $f_{\rm H_2}$
($f_{\rm H_2}$, lower) during $1.1$ Gyr dynamical evolution of disks on $M_{\rm h}$ (left)
and dust-to-gas-ratio 
 $D$ (right) in the selected 23 models. The models with no ${\rm H_2}$ formation are plotted
as $\log M_{\rm H_2}=3$ and $\log f_{\rm H_2}=-4$ just for convenience. 
It should be noted that some low-mass dwarf models with $M_{\rm h}\le 10^{10} {\rm M}_{\odot}$
have no ${\rm H_2}$ (thus no SF), which means that these disks can be identified as `H~{\sc i} dark
galaxies'.
}
\label{Figure. 16}
\end{figure*}

\subsection{Parameter dependences}

The physical properties of ${\rm H_2}$ such as radial profiles of
$R_{\rm mol}$ and time evolution of $f_{\rm H_2}$ 
can be quite diverse in the simulated disk galaxies with different
model parameters. Since it is not so meaningful 
(and much less productive) to describe the results of
all models in detail,  we here
describe only the results that are quite important and thus worth
mentioning. 
In the following subsections,  ${\rm H_2}$ properties (e.g.,  $f_{\rm H_2}$)  are 
compared between models with different values of a model
parameter (e.g., $f_{\rm g}$) for a given set of other model parameters
(i.e., other model parameters are fixed).

\subsubsection{${\rm H_2}$ morphology}

Although the morphologies of ${\rm H_2}$ distributions in disk galaxies
depend largely on the adopted $f_{\rm b}$ and $f_{\rm g}$ in disk galaxies,
almost all models in the present study shows 
very strong central concentration of ${\rm H_2}$ 
and clear coincidence between the locations of gas particles
with high ${\rm H_2}$ fractions and those of spiral arms in the disks.
The distributions of ${\rm H_2}$ along gaseous spiral arms appear to be
very clumpy, which reflects the fact that only the high-density parts of
the spiral arms can efficiently form ${\rm H_2}$. These clumpy distributions
are in a striking contrast with the H~{\sc i} distributions in disk galaxies.

Furthermore, the detailed spatial distributions of ${\rm H_2}$
in the central few kpc
of the disks can be controlled by the presence (or the absence) of
stellar bars developed from bar instability in the central regions.
Fig. 10 shows one of clear examples of the strong influences of the central
stellar bars on the spatial distributions of ${\rm H_2}$ in disk galaxies.
Clearly, the disk galaxy in this model (M8) has a very elongated ring-like structure
dominated by ${\rm H_2}$, which is formed as a result of dynamical action
of the central bar on the gas disk. This kind of `molecular gas ring' can be
seen in the models in which $f_{\rm b}$ is not so large and thus  stellar bar formation
can not be severely suppressed by the central big bulges.

\subsubsection{Time evolution of  $f_{\rm H_2}$}

The time evolution of $f_{\rm H_2}$ depend strongly
on $f_{\rm g}$, $f_{\rm bary}$, $f_{\rm b}$, $M_{\rm h}$,
and the formation redshifts of galaxies in
the present models. Furthermore,  tidal galaxy interaction can significantly
change $M_{\rm H_2}$ and $f_{\rm H_2}$ depending on the orbit configurations
of interacting two galaxies.  
The time evolution of 2D $\Sigma_{\rm HI}$ and $\Sigma_{\rm H_2}$ distributions for
interacting galaxies is
discussed in Appendix B, because the formation process of ${\rm H_2}$
is quite interesting for interacting galaxies. The final 
2D $\Sigma_{\rm HI}$ and $\Sigma_{\rm H_2}$ distributions
for a  high-$z$ disk model are in a striking contrast with those of low-$z$ disks
and thus shown in Appendix B.

Since the time evolution
of $M_{\rm H_2}$ and $f_{\rm H_2}$ are very similar in each model,
we briefly summarize
the derived key 7  dependences of $f_{\rm H_2}$ (not $M_{\rm H_2}$)
on the model parameters in Figs. 11 and 12.
Firstly,  $f_{\rm H_2}$ can be initially larger in 
the models with larger $f_{\rm g}$ for a given set of other model parameters
(e.g., $f_{\rm b}$). The final values of these two quantities ($T=1$ Gyr)
in the model with larger $f_{\rm g}$ (=0.27), 
however, can become rather small owing to rapid ${\rm H_2}$ gas consumption
by star formation. As a result of this, the model with larger $f_{\rm g}$
can have smaller final $M_{\rm H_2}$ and $f_{\rm H_2}$ than
the model with smaller $f_{\rm g}$ (=0.09) . It should be noted that the final 
$f_{\rm g}$ in the model with $f_{\rm g}=0.27$ can become rather low too.

Secondly, $f_{\rm H_2}$ can be initially  smaller in the models
with smaller $f_{\rm bary}$ for a given $f_{\rm g}$ (and $M_{\rm h}$), 
mainly  because formation of non-axisymmetric structures
such as bars and spirals can be suppressed in less strongly self-gravitating
disks (owing to the dominant dark matter halos) so that ${\rm H_2}$ formation
efficiency can be lower. The lower surface gas densities in the models with
smaller $f_{\rm bary}$ are also responsible for the smaller $f_{\rm H_2}$. 
The smaller $f_{\rm H_2}$ for smaller $f_{\rm bary}$ can be seen
in  low-mass models with $M_{\rm h}=10^{10} {\rm M}_{\odot}$.
As shown in the model with smaller $f_{\rm bary}$ (0.02) yet larger $f_{\rm g}$ (=0.33),
$f_{\rm H_2}$ can be larger than the model with $f_{\rm bary}=0.03$ and smaller $f_{\rm g}$.
Since gas consumption by star formation is slower in these models with smaller $f_{\rm bary}$,
the initial $f_{\rm H_2}$ difference in the models (i.e., larger $f_{\rm H_2}$ for larger
$f_{\rm g}$) can last  for a long timescale.

Thirdly, $f_{\rm H_2}$ can be {\it initially} 
larger in the models
with smaller bulges, because bigger galactic bulges can slow down or suppress
the formation of spiral arms and bars so that ${\rm H_2}$ formation
efficiencies in gas disks can be also severely suppressed by the bulges.
Fig. 12 shows that although  $f_{\rm H_2}$  is initially larger in the model
with $f_{\rm b}=0.17$ (fiducial model with a smaller bulge)
than in the model with $f_{\rm b}=1$ ($T<0.2$ Gyr), the $f_{\rm H_2}$
difference becomes smaller as the time passes by. After the formation of a stronger
bar in the central region of the fiducial model, the bar dynamically acts on the gas disk
so that a larger amount of gas can be transferred to the central region, where
${\rm H_2}$ formation is enhanced. Consequently, $f_{\rm H_2}$ can become significantly
larger at $T>1.5$ Gyr. This kind of $f_{\rm H_2}$ enhancement by the dynamical action
of a bar can occur to a lesser extend in the big bulge model with $f_{\rm b}=1$ in which
only a weak stellar bar can be formed.

It should be stressed here that the models (M9 and 12)  with rather large $f_{\rm b}$ (2 and 4),
initial $f_{\rm H_2}$ is rather low ($<0.05$). However, the ${\rm H_2}$ consumption by star formation
is much less efficient in the models so that $f_{\rm H_2}$ can keep slowly rising (owing to the higher
gas densities). As a result of this,  these models can finally show larger $f_{\rm H_2}$ ($>0.1$) at $T=1.1$ Gyr.
Therefore, we can not claim that disks with bigger bulges show smaller $f_{\rm H_2}$ in the present study.
Fourthly,  low surface brightness (LSB) disk galaxies can have initially
smaller $M_{\rm H_2}$ and $f_{\rm H_2}$ owing to their low gas densities.
It should be stressed that owing to much less rapid gas consumption
and chemical enrichment, the LSB model can finally have
$M_{\rm H_2}$ and $f_{\rm H_2}$ similar to those of the fiducial model.

Fifthly, galaxy interaction between two disk galaxies can significantly
increase $f_{\rm H_2}$ after the pericenter passage.
This  enhancement of ${\rm H_2}$ formation in disks can be more clearly
seen in the `prograde' 
tidal interaction model (PR)  in which the spin axes of two disks
are roughly parallel to those of their orbital angular momentum.
This  enhancement of ${\rm H_2}$ formation, however, can not be clearly
seen in other interaction models (RE or distant encounter models),
which implies that not all of interacting galaxies show significantly
large $f_{\rm H_2}$ in comparison with isolated field disk galaxies.
Galaxy interaction can also change the distribution of ${\rm H_2}$ and the MCMF,
in particular, for the central region. The MCMFs of interaction models are
discussed in Appendix A.

Sixthly, less massive disk galaxies can have smaller
$f_{\rm H_2}$, firstly because 
the formation of bars and spirals arms can be severely suppressed
in the galaxies (owing to the stronger concentration of dark matter
and smaller $f_{\rm bary}$), and secondly 
because the dust-to-gas ratios ($D$),
which are key factors for ${\rm H_2}$ formation models of the present study,
can be lower in the disks. 
The chemical enrichment processes by SNe and AGB stars
in these low-mass models are so slow (owing to low SFRs) that
$D$ can not increase during the dynamical evolution of the disks. Therefore,
the formation efficiencies of $f_{\rm H_2}$ can be kept lower in these models.
The low-mass model 
with $M_{\rm h}=10^{10} {\rm M}_{\odot}$ and smaller $f_{\rm bary}=0.01$ 
shows rather small $f_{\rm H_2}$, which means that this low-mass disk
galaxy is almost ${\rm H_2}$-less.
The baryonic fraction ($f_{\rm bary}$) is particularly important
in determining the time evolution of $f_{\rm H_2}$ for low-mass disk
galaxies.

Seventhly, disk galaxies at higher $z$ can have larger
$f_{\rm H_2}$ than those at $z=0$, mainly because
disks are more compact so that a larger amount of ${\rm H_2}$ gas
can be formed in the high-density gaseous regions of the disks.
It should be noted that (i) the simulated  MW-type
disk with low $f_{\rm g}=0.09$ at $z=2$ can have a larger $f_{\rm H_2}$
($\sim 0.32$) and (ii) even the less massive disk with 
$M_{\rm h}=10^{11} {\rm M}_{\odot}$ can have a larger 
$f_{\rm H_2}$ ($\sim 0.16$).
These results  imply that high-$z$ disk galaxies are highly
likely to have systematically rather  high ${\rm H_2}$ gas fractions.
The dependences of 
$M_{\rm H_2}$ and $f_{\rm H_2}$  on $f_{\rm g}$
and $M_{\rm h}$ can be seen among the simulated disk galaxies
at $z=2$.

\subsubsection{Radial density profiles of H~{\sc i} and ${\rm H_2}$}

Fig. 13 shows that the five MW-type disk models with different
$f_{\rm g}$, $f_{\rm bary}$,
$f_{\rm b}$, and surface stellar densities have qualitatively
similar projected radial profiles of H~{\sc i}, ${\rm H_2}$, and $R_{\rm mol}$,
though the absolute magnitudes of these surface mass densities are different
between the models. The $\Sigma_{\rm H_2}$ profiles are steeper than
$\Sigma_{\rm HI}$ in the five models,  and the radial profiles
of $R_{\rm mol}$ ($=\Sigma_{\rm H_2}/\Sigma_{\rm HI}$) have negative
slopes (i.e., higher $R_{\rm mol}$ at smaller $R$). The MW-type disk models
with different $f_{\rm b}$ have the steeper slopes  
approximated by $R^{-1.5}$ whereas the model with higher $f_{\rm g}$ (=0.55)
and smaller $f_{\rm bary}=0.03$
and the LSB model have shallower $R_{\rm mol}$ profiles.
The lower $\Sigma_{\rm H_2}$ in the model with larger $f_{\rm g}$ is due
largely to rapid ${\rm H_2}$ consumption in the inner disk.

Fig. 13 furthermore shows that tidal galaxy interaction can increase both
$\Sigma_{\rm HI}$ and $\Sigma_{\rm H_2}$ significantly owing to the strong
gaseous concentration formed in the central regions of the interacting disks
and the tidal arms.
The strong central  ${\rm H_2}$ concentration is due largely to efficient gas-transfer to the
nuclear region induced by tidal bar and gaseous dissipation.
The formation of ${\rm H_2}$ is more efficient in the central region and tidal arms
during galaxy interaction, which is discussed in more detail  in Appendix B.
The radial profile of $R_{\rm mol}$ in
the interaction model at 
$R<1$ kpc is rather flat and there is the lack of ${\rm H_2}$ gas in 
the circumnuclear region ($R \sim 2$ kpc) in the model.

The less massive disk models with $M_{\rm h}=10^{10} {\rm M}_{\odot}$
and $10^{11} {\rm M}_{\odot}$ have shallower $\Sigma_{\rm H_2}$ profiles
thus shallower $R_{\rm mol}$ profiles too, which is in a striking contrast
with the MW-type disk models. 
The origin of the low $R_{\rm mol}$ in low-mass disk models is closely associated
with weak/no spiral and bar structures (due to the  less strongly gravitation disks)
and initially low $D$.
Fig. 14 clearly demonstrates that if both $f_{\rm g}$ and $f_{\rm bary}$ are the same
between the fiducial MW-type disk model and the low-mass dwarf disk model (M25 with
$M_{\rm h}=10^{10} {\rm M}_{\odot}$),  then $\Sigma_{\rm H_2}$ is dramatically
different between the two models.
The maximum $\log \Sigma_{\rm H_2}$ is only $\sim -3.2$ ${\rm M}_{\odot}$ pc$^{-2}$  in the low-mass disk
model,  which means that even if baryonic fractions are rather high,
gas-poor low-mass disk galaxies are unlikely to form ${\rm H_2}$ efficiently.
A more gas-rich dwarf disk model shows a higher $\Sigma_{\rm H_2}$ along S-shaped
gaseous region, but $\Sigma_{\rm H_2}$ is much lower than the MW-type disk model.
The high-z disk models ($z=2$) also have shallower $R_{\rm mol}$ profiles
in the present study.

As shown in preceding sections,
a strong central concentration of ${\rm H_2}$ can occur in
the barred disk galaxies owing to the dynamical action of stellar bars on gas.
After the rapid consumption of ${\rm H_2}$ gas in the centrally concentrated
gas in the barred disk galaxies,  there can be a deficiency of ${\rm H_2}$ in
their central regions. It is confirmed that  the bulge-less disk galaxy
with $f_{\rm b}=0$ at $T=2$ Gyr (i.e., after long-term evolution), 
for which a strong bar can be spontaneously
formed owing to global
bar instability, can finally show the lack of ${\rm H_2}$ in its central region
after rapid ${\rm H_2}$  consumption by star formation  in the central region.
This result implies that if gas accretion onto  disks 
is truncated,  even the late-type
disk galaxies with no/little bulges can show the lack of ${\rm H_2}$ gas
in their central regions. 

\subsubsection{Internal ${\rm H_2}$-scaling relations}

Fig. 15 describes the physical correlations of local properties 
of ${\rm H_2}$ with those of  dust and stars within galaxies with different model
parameters. These `internal' ${\rm H_2}$-scaling relations will be able
to be  compared with ongoing observational studies on ${\rm H_2}$ properties
of disk galaxies (by, e.g., ALMA), and thus the predicted scaling relations
will be useful for interpreting the observational results. It is clear
that the models with different $f_{\rm g}$, $f_{\rm b}$, and initial stellar
mass densities of the disks have a similar positive correlation
between $M_{\rm H_2}/M_{\ast}$ and $\Sigma_{\ast}$ 
for $\log \Sigma_{\ast}<8$ ${\rm M}_{\odot}$ kpc$^{-2}$.
The simulated relatively flat correlation 
for $\log \Sigma_{\ast}\ge8$ ${\rm M}_{\odot}$ kpc$^{-2}$
is due to the strong concentration of ${\rm H_2}$ 
in the central regions (where $\Sigma_{\ast}$ is high) caused
by dynamical action of the stellar bars on gas in the disks.

The tidal interaction model, in which $f_{\rm H_2}$ can be 
enhanced and ${\rm H_2}$ distribution can be significantly changed,
shows a $M_{\rm H_2}/M_{\ast}-\Sigma_{\ast}$ relation similar to
those seen in the isolated MW-type disk models, though the correlation
is not so clear owing to the disturbed ${\rm H_2}$ distribution.
The less massive disk models with low $M_{\rm h}$ also show strong positive 
$M_{\rm H_2}/M_{\ast}-\Sigma_{\ast}$ correlations for lower
$\Sigma_{\ast}$, though the central ${\rm H_2}$ concentration makes
the correlation shallower for higher $\Sigma_{\ast}$.
The high-$z$ models with $z=2$ also have clear positive
$M_{\rm H_2}/M_{\ast}-\Sigma_{\ast}$ correlations, which implies
that a linear $M_{\rm H_2}/M_{\ast}-\Sigma_{\ast}$
correlation will be identified in future observations of ${\rm H_2}$
distributions in high$-z$ disk galaxies.

These ten models show a much tighter correlation between
$\Sigma_{\rm H_2}$ and $\Sigma_{\rm dust}$, which means
that local ISM with higher dust surface densities can have
higher ${\rm H_2}$ surface densities in disk galaxies. 
This simulated almost linear 
$\Sigma_{\rm H_2}-\Sigma_{\rm dust}$ scaling relation is not
so surprising, because ${\rm H_2}$ formation efficiencies in local ISM
are assumed to depend on the dust-to-gas ratios of local ISM in
the present study. 
Recent observational studies have just started a detailed investigation
on the sub-kpc-scale properties of dust in nearby disk galaxies
like M31 (e.g., Viaene et al. 2014), which can be combined with
the observed sub-kpc-scale properties of ${\rm H_2}$ to produce a possible
dust-${\rm H_2}$ scaling relation within galaxies.
The observed $\Sigma_{\rm H_2}-\Sigma_{\rm dust}$ scaling relation within
galaxies will be able to be used as a stringent test for any theoretical
model of ${\rm H_2}$ formation on dust grains in galaxies.

\subsubsection{Maximum possible ${\rm H_2}$ mass}

Fig. 16 describes the dependence of the maximum values
of $M_{\rm H_2}$ and $f_{\rm H_2}$ during $\sim 1$ Gyr dynamical evolution
on the initial halo masses ($M_{\rm h}$) and dust-to-gas-ratios ($D$)
in the selected models.
It is clear from this figure that the models with larger $M_{\rm h}$
or larger $D$ 
are more likely to have larger $M_{\rm H_2,max}$ and $f_{\rm H_2,max}$
The low-mass models have initially smaller $D$ due to
the adopted mass-metallicity relation,
and spiral arms and bars,
which are the major driver for efficient ${\rm H_2}$ formation in disks,
can not be formed owing to the less strongly self-gravitating disks
(i.e., more strongly dominated by dark matter). This combination
of the low $D$ and the incapability of bar/spiral  formation in the low-mass models
is the main reason for the derived dependences of 
$M_{\rm H_2,max}$ and $f_{\rm H_2,max}$ on $M_{\rm h}$ and $D$.

It should be stressed here that some models with 
$M_{\rm h} \le 10^{10} {\rm M}_{\odot}$
and those with $\log D <-4$ have no ${\rm H_2}$ in the disks
and thus no star formation  in the present star formation  models in which
new stars can form exclusively from ${\rm H_2}$ gas.
This means that low-mass disk galaxies with very low $D$ are dominated
by H~{\sc i} gas and have no star formation so that they can be identified
as isolated massive H~{\sc i} clouds with no star formation
(in other words, 'dark H~{\sc i} galaxies').
The total ${\rm H_2}$ mass can not become larger than $10^6 {\rm M}_{\odot}$ in
some of low-mass models with $M_{\rm h}<10^{10} {\rm M}_{\odot}$,
which means that  massive star clusters and globular clusters 
are unlikely to
be formed from GMCs in low-mass disk galaxies. We will discuss this point
later in the Discussion section.

\section{Discussion}


\subsection{What determines $f_{\rm H_2}$ ($R_{\rm mol}$) ?}

One of the unresolved problems related to ${\rm H_2}$ contents of galaxies
dependent on Hubble types is that 
Sc/Sd late-type disks have   smaller $f_{\rm H_2}$ ($R_{\rm mol}$) in spite of their having
larger gas masses with respect to their dynamical masses ($M_{\rm gas}/M_{\rm dyn}$, see Figs. 4 and 5
in YS91, as an example for this).
Although this problem will be able to be better discussed in our forthcoming papers in which
$\Lambda$ CDM galaxy formation  models with ${\rm H_2}$ formation  are adopted,
we can discuss this problem  based on the present new results.
Galactic  bulges have been demonstrated to suppress ${\rm H_2}$ formation in ISM
only for  the early dynamical evolution of disk galaxies,
if $f_{\rm b}$ is quite large ($1-4$, corresponding to Sa/S0 galaxies).
The main reason for this is that
spiral arms and bars, which can enhance the ${\rm H_2}$ formation,
can be suppressed by the presence of big bulges.
The stronger ISRFs of big and dense bulges can be also responsible for less
efficient conversion from H~{\sc i} to ${\rm H_2}$ in ISM.
Such a suppression effect of ${\rm H_2}$ formation can be seen in models 
with different $M_{\rm h}$ and $f_{\rm g}$.

However, the larger $f_{\rm H_2}$ in disks with smaller bulges can not last long owing
to the more rapid gas consumption by star formation. Furthermore,  big bulge can severely suppress
the star formation so that $f_{\rm H_2}$ in disks with big bulges
can steadily increase owing to chemical enrichment (i.e., increasing $D$)
and finally become significantly larger.
This means that although $f_{\rm H_2}$ can be larger in disks with smaller bulges
{\it for a fixed} $f_{\rm g}$,  such disks do not necessarily continue to show larger $f_{\rm H_2}$
(particularly when there is no external gas supply).
Also the initially larger $f_{\rm H_2}$ in disks with smaller bulges appears to be inconsistent with
the observed trend of $f_{\rm H_2}$ with Hubble types.
Given these results, it appears difficult to understand
the origin of smaller $f_{\rm H_2}$ in late-type Sc/Sd galaxies 
in the context of the bulge effects on ${\rm H_2}$ formation.

The central stellar bars formed from global bar instability have been demonstrated to enhance
significantly $f_{\rm H_2}$ ($R_{\rm mol}$), in particular, in the central regions of disk
galaxies, though such enhancement can not last long owing to rapid gas consumption by
star formation. Strong spirals arms have been demonstrated to be 
where ${\rm H_2}$ formation is very efficient in the present study.
Given that global bar/spiral  instability can be prohibited by big bulges,
disks with later types (i.e., smaller bulges) are more likely to have bars and thus show
larger $f_{\rm H_2}$. This possible trend of $f_{\rm H_2}$ with Hubble types is opposite
to the observed trend, which implies  that the formation efficiency of stellar bars is not
the major factor that determines $f_{\rm H_2}$ in Sc/Sd galaxies.

The present study has also shown that the final $f_{\rm H_2}$ of disk galaxies
depend on the initial baryonic mass fractions ($f_{\rm bary}$) such that
$f_{\rm H_2}$ can be larger  in the models with  larger  $f_{\rm bary}$
for a given $f_{\rm g}$ ($M_{\rm h}$).
Recent observations have shown that $f_{\rm bary}$ can be smaller in galaxies
with lower halo masses (e.g., Papastergis et al. 2012).
These observations combined with 
the present results therefore imply that
if the late-type Sc/Sd galaxies can be preferentially formed from 
dark halos with lower $M_{\rm h}$, then the observed systematically small $f_{\rm H_2}$
can be more clearly explained. Although the observed luminosity functions
for different Hubble types in the local field and the  Virgo cluster suggests that
Sa+Sb galaxies are systematically brighter than  Sc+Sd galaxies 
(e.g., Binggeli et al. 1988),  observational studies 
have not yet determined the mass functions of dark matter halos and baryonic fractions for 
different Hubble types. Therefore, it is unclear whether the above explanation
for the observed small  $f_{\rm H_2}$ in Sc/Sd galaxies is really plausible and realistic,
though the smaller  $f_{\rm bary}$ in later Hubble types would be a promising scenario.

\subsection{Do dark H~{\sc i} galaxies exist ?}

The present study has shown that low-mass disk galaxies show rather small 
$f_{\rm H_2, max}$ (maximum possible ${\rm H_2}$ fraction) and therefore
can be identified as  H~{\sc i}-dominated galaxies.
The major physical reason for this very low ${\rm H_2}$ contents is
that these low-mass galaxies are extremely dust-poor so that they
can not form ${\rm H_2}$ on dust grains.
The possible smaller $f_{\rm bary}$ in these low-mass galaxies can also contribute to
the very low ${\rm H_2}$ formation efficiency of the ISM.
Furthermore,
some of the simulated low-mass galaxies with rather
high initial $f_{\rm g}$ ($>0.9$) can show virtually $f_{\rm H_2,max}=0$, 
if $M_{\rm h}$ is as low as or lower than $3 \times 10^9 {\rm M}_{\odot}$. 
These results imply that there would exist low-mass, and extremely low surface-brightness
galaxies with only H~{\sc i} gas: These are almost `dark' galaxies. 

So far no observational studies have found  strong evidence for the presence
of {\it isolated} (or `intergalactic')  dark
H~{\sc i}-rich galaxies  with total gas masses larger than $10^6 {\rm M}_{\odot}$
in the nearby universe, though a number of observational
studies investigated H~{\sc i} contents of galaxies in group environments
(e.g., Kilborn et al 2005;  Pisano et al. 2011). 
Although most of the detected H~{\sc i} objects have optical counterparts
(or tidal origin) in these observations,
a few observations so far discovered massive apparently isolated H~{\sc i} gas
clouds with masses larger than $10^8 {\rm M}_{\odot}$ (e.g., Davies et al. 2004;
Kolibalski et al. 2004).
These intriguing massive isolated objects are likely to be
just tidal debris rather than  dark H~{\sc i} galaxies (e.g., Bekki et al. 2005).
If H~{\sc i}-dominated low-mass dark galaxies do not exist,
the present study implies  that such low-mass galaxies 
had lost almost all of their gas at their formation epochs or during their evolution 
owing to some physical processes such as ram pressure stripping and cosmic reionization.

\subsection{Implications on the formation of  globular clusters at high $z$}

The present study has shown that (i) some low-mass galaxies
with $M_{\rm h} \le 10^{10} {\rm M}_{\odot}$  can not form ${\rm H_2}$ efficiently
owing to their low dust masses  and consequently (ii) the total 
${\rm H_2}$ masses are unlikely to exceed $10^7 {\rm M}_{\odot}$. These results
can have the following implications on the formation of massive star clusters,
in particular,  globular clusters (GCs),  the origin of which remains unexplained.
As recent theoretical studies of GC formation have demonstrated
(e.g., D'Ercole et al. 2008; Bekki 2011),
the initial stellar masses of GCs should be $\sim 10$ times larger than the present
ones (typically, $2 \times 10^5 {\rm M}_{\odot}$): Here we do not discuss
the physical reasons for the proposed rather high initial masses of GC progenitor.
This means that the original GMCs from which the initial massive stellar systems
can form
should be larger than $\sim 10^7 {\rm M}_{\odot}$ even for
a high star formation efficiency of $\sim 0.2$.

These theoretical results on GC formation combined with the present results imply that
low-mass galaxies with $M_{\rm h} \le 10^{10} {\rm M}_{\odot}$ 
are unlikely to form GCs within them owing to the low ${\rm H_2}$ 
contents. It would be possible that only low-mass, low-density star clusters,
can form in their host low-mass galaxies.
These low-mass clusters are prone to tidal destruction by their host galaxies
so that they can finally become field stars within their hosts.
If there exists a threshold halo mass above which GCs can form, then
it can not only explain the observed absence of GCs in some faint dwarf galaxies
in the Local Group (e.g., van den Bergh 2000) but also provide a physical basis
on the minimum halo mass of GC host galaxies introduced in recent semi-analytic
models of GC formation in galaxies 
based on hierarchical galaxy formation scenarios
(e.g., Bekki et al. 2008; Griffen et al. 2010).

\section{Conclusions}

We have investigated the physical properties of ${\rm H_2}$ in disk galaxies with different
masses and Hubble types based on our new model for the formation of ${\rm H_2}$ on dust grains.
The basic parameters in this numerical study are $M_{\rm h}$ (halo mass), $f_{\rm bary}$
(baryonic mass fraction), $f_{\rm g}$ (gas mass fraction),
and $f_{\rm b}$ (bulge-to-disk-ratio). We have investigated how the spatial distributions,
time evolution, and scaling relations of ${\rm H_2}$ depend on these parameters both for isolated
disk galaxies and for interacting ones. The principal results are as follows. \\

(1) The observed positive correlation between the mass ratio of ${\rm H_2}$ to
H~{\sc i} ($R_{\rm mol}$) 
and gaseous pressure ($P_{\rm g}$) can be reproduced reasonably well
by the present models. However, the $R_{\rm mol}-P_{\rm g}$ relation
($R_{\rm mol} \propto P_{\rm g}^{0.92}$) can be reproduced well
only for some ranges of  model parameters for dust growth and destruction.
These results imply that dust can play a significant role in the formation
of the $R_{\rm mol}-P_{\rm g}$ relation
and thus that dust formation and evolution needs to be included
in discussing the origin of ${\rm H_2}$ scaling relations. \\

(2) The simulated ${\rm H_2}$ surface density distributions (H2SDDs)
in isolated luminous disk galaxies  can have  the slope of $\sim -1.5$ 
for $\log \Sigma_{\rm H_2} < 2.4$ ${\rm M}_{\odot}$ pc$^{-2}$.
The simulated slope  is similar to the slope of the observed molecular cloud
mass function for GMCs in the MW ($N_{\rm mc} \propto m_{\rm mc}^{-1.5}$).
Although the simulated slopes of H2SDDs does not depend so strongly on
galaxy parameters such as $f_{\rm b}$ and  $f_{\rm gas}$ in isolated models,
the mean $\Sigma_{\rm H_2}$  can be significantly increased in tidal interaction
models.  \\

(3) The simulated radial profiles of the $\Sigma_{\rm H_2}$-to-
$\Sigma_{\rm HI}$-ratios in isolated luminous disk galaxies
can be described as  $R^{-1.5}$,
which  are observed in disk galaxies.  This successful reproduction
of observations implies that the present dust-regulated ${\rm H_2}$ models
can grasp some essential ingredients of ${\rm H_2}$ formation in ISM of disk galaxies.
The radial gradients can be different between low-mass and high-mass disk galaxies
at low and high $z$ and between isolated and interacting galaxies. 
The simulated ${\rm H_2}$ distributions are clumpier than H~{\sc i}
in most models. The ${\rm H_2}$ surface density of a disk galaxy can dramatically
drop beyond a certain radius where ${\rm H_2}$ formation efficiency becomes rather
low owing to the low gas density and the low $D$.
\\

(4) The mass-ratios of ${\rm H_2}$ to stars ($M_{\rm H_2}/M_{\ast}$)
in individual local regions
of a galaxy can anti-correlate with the local surface stellar density ($\Sigma_{\ast}$). 
This anti-correlation ($M_{\rm H_2}/M_{\ast} \propto \Sigma_{\ast}^{-1}$) 
can be seen in different galaxy  models with different parameters.
The ${\rm H_2}$ surface densities ($\Sigma_{\rm H_2}$)
of local individual regions in a galaxy
can correlate with the dust surface densities ($\Sigma_{\rm dust}$) such that
$\Sigma_{\rm H_2} \propto \Sigma_{\rm dust}$.
This ${\rm H_2}$-dust correlation is rather tight in different galaxy models. \\

(5) The initial total masses of dark halos ($M_{\rm h}$) can control the
${\rm H_2}$ mass fractions ($f_{\rm H_2}$) of disk galaxies  in such a way that 
$f_{\rm H_2}$ can be larger for larger $M_{\rm h}$.
This is firstly because the formation of spiral arms and bars, where ${\rm H_2}$ formation
is rather efficient, can be severely suppressed in smaller $M_{\rm h}$,
and secondly because the dust-to-gas-ratios ($D$), which determines the formation
efficiency of ${\rm H_2}$ on dust grains, are lower for smaller $M_{\rm h}$. 
Some low-mass galaxies with $M_{\rm h} \le 10^{10} {\rm M}_{\odot}$ show no/little ${\rm H_2}$
formation within their disks in the present study. \\

(6) Galactic bulges  can strongly suppress the formation of ${\rm H_2}$ in gas disks
at the early dynamical evolution of disk galaxies, if $f_{\rm b}$ is quite large ($\ge 1$).
This is  mainly because
big bulges can prevent global spiral/bar instability from occurring so that
efficient ${\rm H_2}$ formation within spirals and bars can be severely suppressed.
This, however, does not necessarily mean that $f_{\rm H_2}$ can continue to be low,
because slow gas consumption by star formation yet steady chemical enrichment (i.e., increasing $D$) can finally
enhance $f_{\rm H_2}$ for some  disk models with big bulges in the present study.
These results imply that the origin of some late-type disk galaxies  (Sc and Sd) with smaller $f_{\rm H_2}$ 
is not simply due to their small bulges. Some disk models in the present study
show the lack of ${\rm H_2}$ gas
in their central regions. \\

(7) Barred disk galaxies are more likely to have larger $f_{\rm H_2}$ owing to the formation of
centrally concentrated ${\rm H_2}$ rings and disks.  The central ${\rm H_2}$ can be rapidly consumed
by star formation so that higher $f_{\rm H_2}$ in barred galaxies
can not last long without further external gas supply onto the
disks. Some of the simulated barred disk galaxies can have elongated ${\rm H_2}$-rings in their
central regions, though such ${\rm H_2}$ structures are short-lived. 
Bulge-less disk galaxies with central bars can also show the lack of ${\rm H_2}$ in their central regions
after rapid consumption of ${\rm H_2}$ transferred to the inner few kpc.
\\

(8) Disk galaxies at higher $z$ are more likely to have larger $f_{\rm H_2}$ owing to their initially
high gas densities in the more compact disks. This result does not depend on other model parameters
such as $M_{\rm h}$. Disk galaxies with smaller $f_{\rm bary}$ at higher $z$ can still show 
higher $f_{\rm H_2}$. For a given $M_{\rm h}$, $f_{\rm g}$, and $z$,  
disk galaxies with smaller $f_{\rm bary}$ have
smaller $f_{\rm H_2}$.
High-$z$ disks are more likely to have very clumpy ${\rm H_2}$ distributions with numerous
small ${\rm H_2}$ clouds.

(9) The maximum ${\rm H_2}$ masses ($M_{\rm H_2,max}$) and fractions 
($f_{\rm H_2,max}$) depend on initial halo masses $M_{\rm h}$ and dust-to-gas-ratios
($D$) such that both can be higher in larger $M_{\rm h}$ and larger $D$. 
For low-mass halos with $M_{\rm h} \le 10^{10} {\rm M}_{\odot}$,
the formation of ${\rm H_2}$ can be completely suppressed in some models
with smaller gas fractions so that no stars
can be formed. Some low-mass galaxies with $M_{\rm h} \le 3 \times 10^9 {\rm M}_{\odot}$
can not form ${\rm H_2}$ owing to rather low dust contents even if
they are quite gas-rich $f_{\rm g}>0.9$.
These galaxies with no ${\rm H_2}$ are still very rich in H~{\sc i} 
and could  be observationally identified as gas-rich extremely low-surface brightness
galaxies (or almost 'dark galaxies'). \\

(10) The spatial distributions of ${\rm H_2}$ in the simulated galaxies are quite diverse
depending on initial $M_{\rm h}$ and Hubble types (e.g., bulge-to-disk-ratio, $f_{\rm b}$).
Circumnuclear ${\rm H_2}$ rings can be formed in barred disk galaxies,
and such ring formation can significantly increase $f_{\rm H_2}$ in the galaxies.
Radial ${\rm H_2}$ profiles can be steeper for more luminous disk galaxies in isolation,
and tidal interaction can dramatically increase the degrees of central ${\rm H_2}$ 
concentration in disk galaxies. \\

(11) The present results have several important implications of star formation processes
of galaxies. For example, massive star clusters, like old globular clusters,
are unlikely to be formed in low-mass galaxies
($M_{\rm h} < 3 \times 10^{9} {\rm M}_{\odot}$),  because the maximum possible total
${\rm H_2}$ masses  can be well less than $10^6 {\rm M}_{\odot}$ (i.e.,
owing to the incapability of giant molecular clouds to form).
This result implies that there would exist a threshold galaxy mass above which
old GCs can be formed within galaxies at high $z$.\\

\section{Acknowledgment}
I (Kenji Bekki; KB) am   grateful to the referee  for  
constructive and useful comments that improved this paper.
Numerical simulations  reported here were carried out on
the three GPU clusters,  Pleiades, Fornax,
and gSTAR kindly made available by International Center for radio astronomy 
research (ICRAR) at  The University of Western Australia,
iVEC,  and the Center for Astrophysics and Supercomputing
in the Swinburne University, respectively.
This research was supported by resources awarded under the Astronomy Australia 
Ltd's ASTAC scheme on Swinburne with support from the Australian government. gSTAR
is funded by Swinburne and the Australian Government's
Education Investment Fund.
KB acknowledges the financial support of the Australian Research Council
throughout the course of this work.

\appendix

\section{The formation of a central stellar bar}

In the present fiducial MW-type disk model (M1), a stellar bar can play a significant role
in the formation of ${\rm H_2}$ on dust grains and the dynamical evolution of ${\rm H_2}$ gas,
in particular, in the later evolution of the central region of the gas disk. 
We here briefly describe   the bar formation process in this model. 
As shown in Fig. A1 on the time evolution of stellar surface density ($\Sigma_{\rm s}$) of the disk,
a stellar bar can start to develop around  $T=0.8$ Gyr owing to bar instability in this model
with a relatively small bulge mass. The formation of  a strong nuclear bar can be completed by
$T=1.1$ Gyr so that the bar can influence the gas dynamics there after its formation
(e.g., a massive nuclear ${\rm H_2}$ disk can be formed at $T=2$ Gyr as a result of dynamical interaction
between the bar and ${\rm H_2}$ gas, as shown in Fig. 4).
The gaseous component (H~{\sc i}+${\rm H_2}$) also shows a bar-like distribution in the central region
of the disk when the stellar bar forms at $T=1.1$ Gyr.

\begin{figure*}
\psfig{file=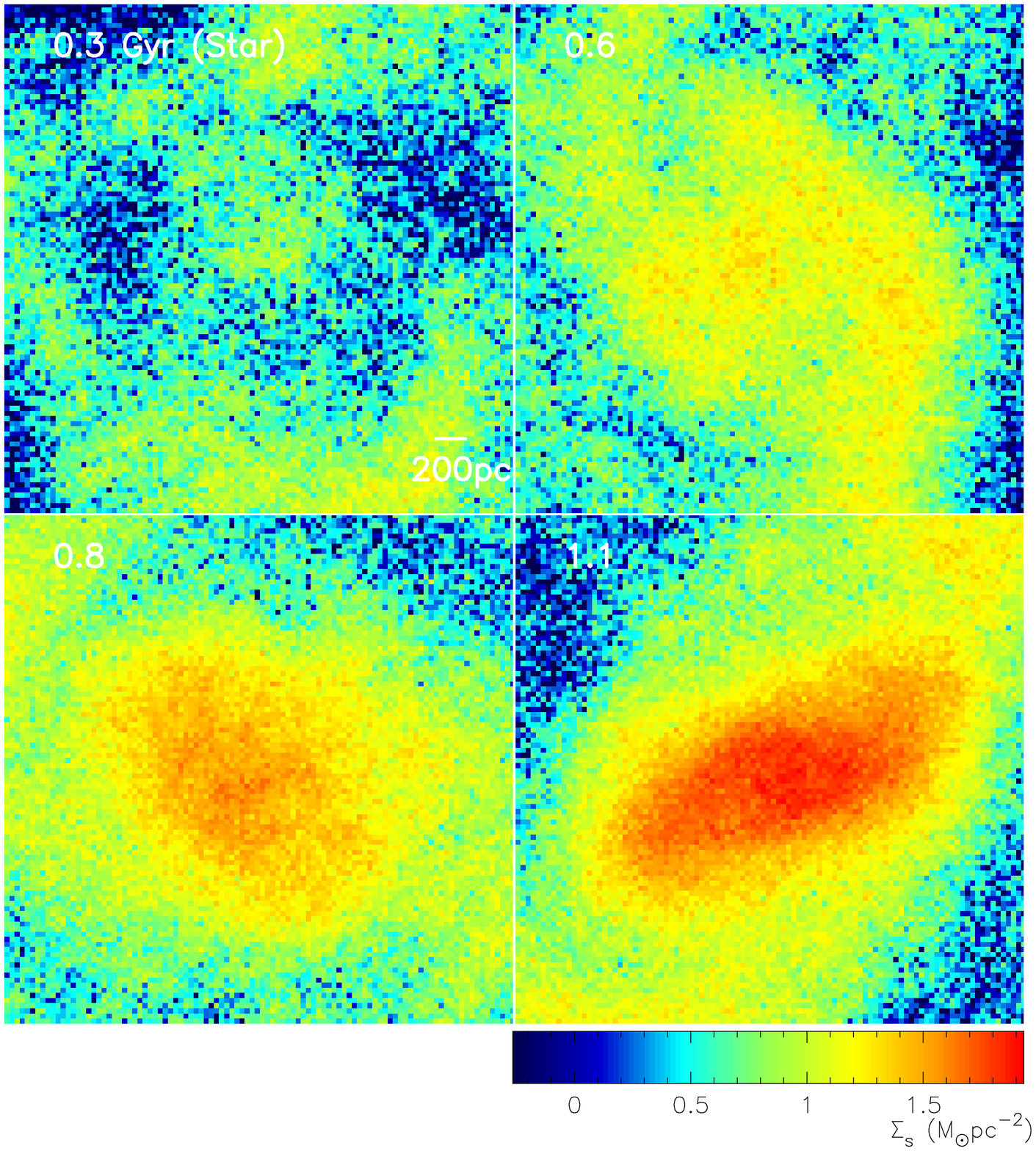,width=8.0cm}
\caption{
The time evolution of the projected mass densities for stars ($\Sigma_{\rm s}$)
in the central region
of the  fiducial MW-type disk model (M1). The formation of a central stellar  bar can be developed
within less than 1 Gyr in this model so that the bar can influence the formation of 
${\rm H_2}$ there.
}
\label{Figure. A1}
\end{figure*}

\section{Intriguing ${\rm H_2}$ distributions in selected   models}

Figs. B1 and B2 show intriguing ${\rm H_2}$ distributions of galactic disks in  selected models.
In Fig. B1, the initial $\Sigma_{\rm H_2}$ of each mesh  at $T=0$
is calculated by assuming that ${\rm H_2}$ mass fraction
is 0.01 just for convenience, because ${\rm H_2}$ formation efficiency based ISRF etc is not calculated at $T=0$.
The formation efficiency of ${\rm H_2}$ on dust grains can be significantly enhanced
in the strong tidal arms of the interacting disk galaxy  for
the  prograde interaction model  (M16),
as shown in Fig. B1 (at $T=0.6$ Gyr). During the efficient gas infall onto the nuclear regions,
both H~{\sc i} and ${\rm H_2}$ gas densities can become significantly higher so that star formation rates 
can be high in the nuclear region and the tidal arms.
After tidal interaction, the ${\rm H_2}$ distribution appears to be clumpy with a number of
off-center high-density ${\rm H_2}$ clumps in the disk.
Such off-center massive gas clumps can be also seen in recent models of gas-rich interacting
galaxies by Yozin \& Bekki (2014b).
 Since the number of tidal interaction
 models investigated in the present study is rather small ($<10$), we can not conclude whether
such a clumpy ${\rm H_2}$ distribution is a characteristic of post-interacting galaxies.
We will investigate this point in our future works by performing a larger number of numerical simulations
of interacting galaxies with a  much wider range of model parameters
for galaxy interaction.

Fig. B2 shows that a gas-rich, more compact disk at $z=2$ can have a widespread spatial distribution
of ${\rm H_2}$ gas in the model M15. In this model, the baryonic mass fraction is lower than the fiducial
MW-disk model (i.e., less strongly self-gravitating) so that the formation of a strong bar can be suppressed.
Therefore, a bar-like distribution of gas can not be clearly seen in this model. Even in the outer
part of the gas disk, high-density ${\rm H_2}$ regions can be seen, which is in a striking contrast with
the ${\rm H_2}$ distribution of the low-$z$ fiducial MW-disk model (M1). The gas disk at $z=2$ appears to
be composed of numerous small ${\rm H_2}$ clumps, which implies that the mass function of molecular clouds
(MCMF)
can be significantly different between $z=0$ and $z=2$. Since the present simulation does not have enough
resolution to investigate MCMFs, we will discuss this point in our future numerical works with a much
larger number of gas particles.

\begin{figure*}
\psfig{file=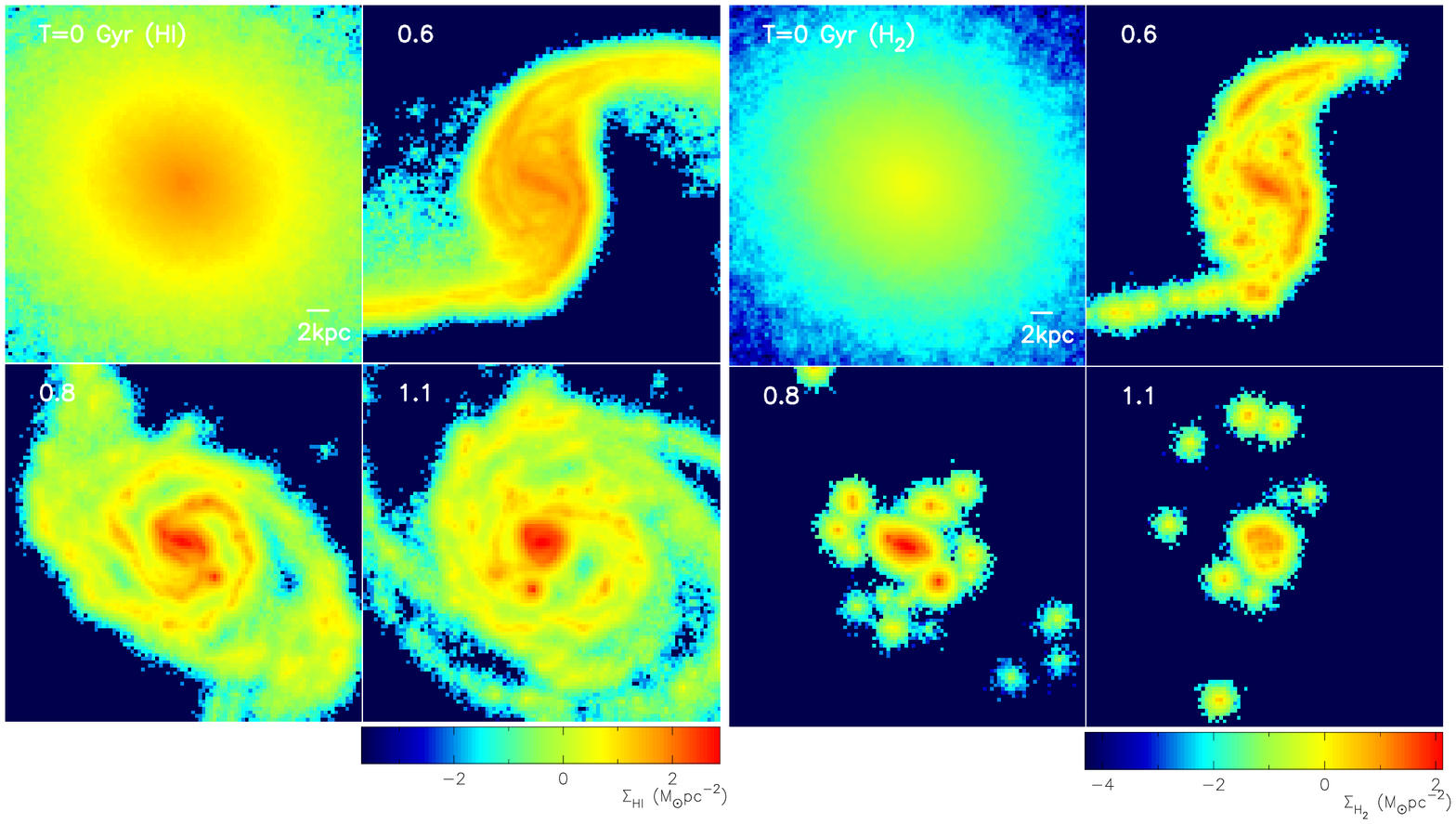,width=18.0cm}
\caption{
The time evolution of the projected mass densities for H~{\sc i} ($\Sigma_{\rm HI}$, left four)
and ${\rm H_2}$  ($\Sigma_{\rm H_2}$, right four) for the prograde tidal interaction model  (M16). 
}
\label{Figure. B1}
\end{figure*}

\begin{figure*}
\psfig{file=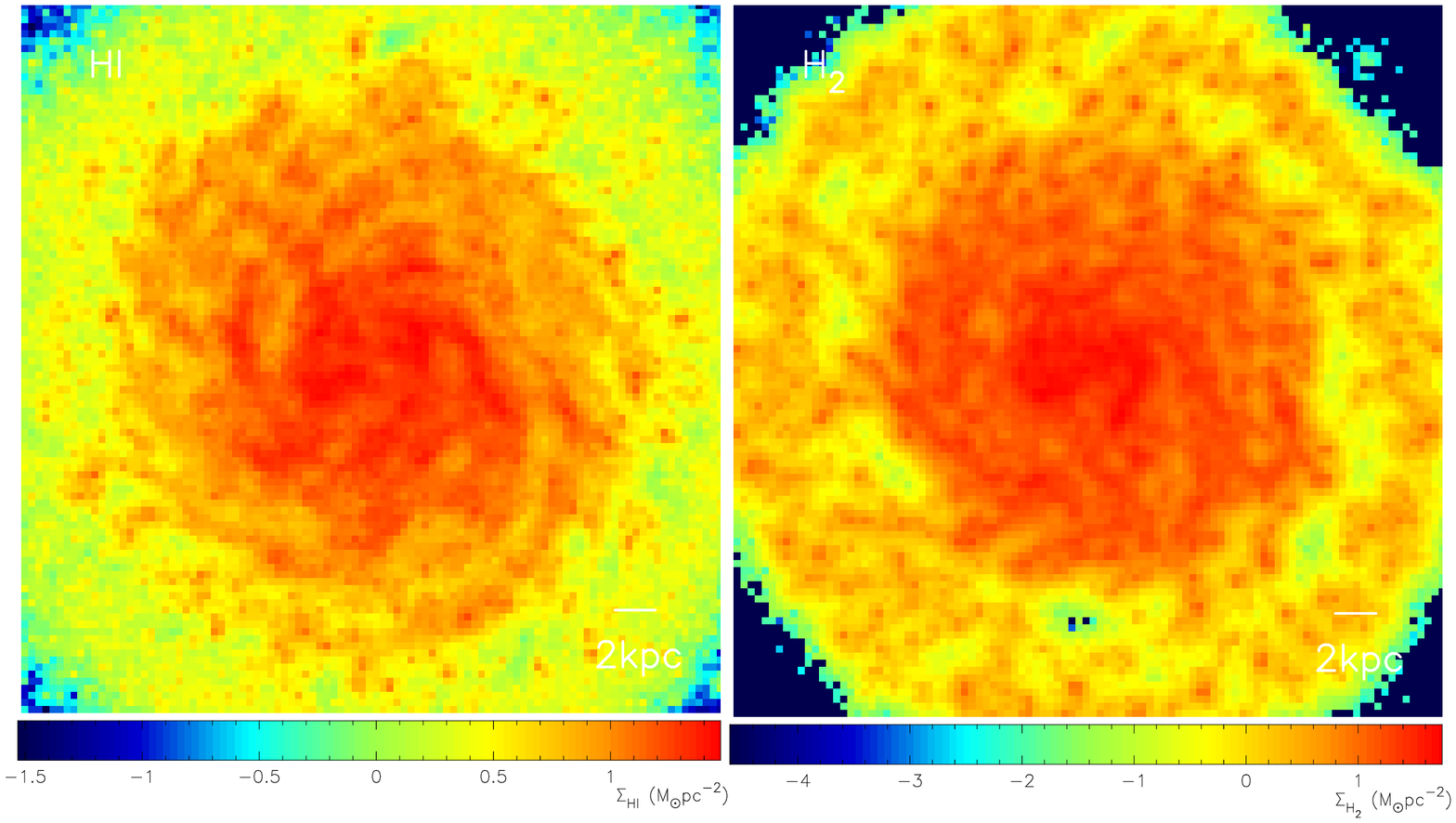,width=9.0cm}
\caption{
The final  projected mass densities for H~{\sc i} ($\Sigma_{\rm HI}$, left)
and ${\rm H_2}$  ($\Sigma_{\rm H_2}$, right) for the high-$z$ MW-type disk model with
$f_{\rm g}=0.55$ (M15).
}
\label{Figure. B2}
\end{figure*}
\end{document}